\documentclass[pss]{wiley2sp} % provides pss two-column style
\usepackage[x11names]{xcolor}
\usepackage{amsmath}
\usepackage{amsmath,amssymb,epic,eepic}
\usepackage{siunitx}                % unités SI
\usepackage{graphicx}
\usepackage{tikz}
  \usetikzlibrary{patterns}
  \usetikzlibrary{positioning}
  \usetikzlibrary{shapes,shapes.misc,shapes.geometric} 
  \usetikzlibrary{arrows}
  \usetikzlibrary{calc}
  \usetikzlibrary{matrix}
  \usetikzlibrary{intersections}
  \usetikzlibrary{decorations,decorations.markings,decorations.pathreplacing,
decorations.pathmorphing}
\usepackage{braket}
\usepackage{bbding}
\usepackage{pifont}

\usepackage{todonotes}

\newcommand{\me}{\mathrm{e}}
\newcommand{\mi}{\mathrm{i}}
\newcommand{\md}{\mathrm{d}}

\colorlet{myred}{Firebrick3}
\colorlet{myblue}{SteelBlue4}
\colorlet{mygreen}{SeaGreen4}
\colorlet{mygray}{Ivory4}
\colorlet{mypurple}{DarkOrchid4}

\begin{document}

\title{Electron quantum optics as quantum signal processing}

\titlerunning{Electron quantum optics as quantum signal processing}
\author{B. Roussel\textsuperscript{\textsf{\bfseries 1}}, 
C. Cabart\textsuperscript{\textsf{\bfseries 1}}, 
G. F\`eve\textsuperscript{\textsf{\bfseries 2}}, 
E. Thibierge\textsuperscript{\textsf{\bfseries 1}}, 
P. Degiovanni\textsuperscript{\textsf{\bfseries 1}}}

%\author{B. Roussel\textsuperscript{\textsf{\bfseries 1}}} 
%\author{C. Cabart\textsuperscript{\textsf{\bfseries 1}}} 
%\author{G. F\`eve\textsuperscript{\textsf{\bfseries 2}}} 
%\author{E. Thibierge\textsuperscript{\textsf{\bfseries 1}}} 
%\author{P. Degiovanni\textsuperscript{\textsf{\bfseries 1}}}
%
%\affiliation{\textsuperscript{1}\,Univ Lyon, Ens de Lyon, Universit\'e Claude
%Bernard Lyon 1, CNRS,\\ 
%Laboratoire de Physique, F-69342 Lyon, France}
%
%\affiliation{\textsuperscript{2}\,Laboratoire Pierre Aigrain, Ecole Normale
%Sup\'erieure-PSL Research University,\\
%CNRS, Universit\'e Pierre et Marie Curie-Sorbonne Universit\'es,
%Universit\'e
%Paris
%Diderot-Sorbonne Paris Cit\'e,\\ 24 rue Lhomond, 75231 Paris Cedex 05,
%France}

\institute{\textsuperscript{1}\,Univ Lyon, Ens de Lyon, Universit\'e Claude
Bernard Lyon 1, CNRS,\\ 
Laboratoire de Physique, F-69342 Lyon, France\\
\textsuperscript{2}\,Laboratoire Pierre Aigrain, Ecole Normale
Sup\'erieure-PSL Research University,\\
CNRS, Universit\'e Pierre et Marie Curie-Sorbonne Universit\'es,
Universit\'e
Paris
Diderot-Sorbonne Paris Cit\'e,\\ 24 rue Lhomond, 75231 Paris Cedex 05,
France}

% Abbreviated list of authors for the page headers
\authorrunning{B. Roussel et al.}

%E-mail-address of corresponding author
\mail{mail: \textsf{Pascal.Degiovanni@ens-lyon.fr}}
%  \textsf{Pascal.Degiovanni@ens-lyon.fr}}

\abstract{
\abstcol{
	The recent developments of electron quantum optics in quantum Hall edge
channels have given us new ways to probe the behavior of electrons in
quantum conductors. 
It has brought new quantities called
electronic coherences under the
spotlight. In this paper, 
we explore the relations between electron quantum optics and
signal processing
through a global review of the various methods for
accessing
single- and two-electron coherences in electron quantum
optics. We interpret
electron quantum optics interference experiments as analog signal processing,
converting quantum signals into experimentally observable quantities
such as current averages and correlations.}{
	This point of view also gives us a procedure
to obtain
quantum information quantities from electron quantum
optics coherences. We illustrate these ideas by discussing two-mode
entanglement in electron quantum optics. We also sketch how
signal processing ideas may open new perspectives for representing 
electronic coherences in
quantum conductors and understand the properties of the underlying many-body
electronic state.
}
% End of abstract...
}

\maketitle

\section{Introduction}
\label{sec:introduction}

Electron quantum optics is a new perspective on electronic 
quantum transport that aims at understanding 
the behavior of electrons in ballistic quantum conductors
using paradigms and methods of quantum optics~\cite{Bocquillon:2014-1}. 
This field has emerged from the
development of quantum coherent nanoelectronics during the
90s~\cite{Henny:1999-1,Oliver:1999-1,Blanter:2000-1}, the
demonstration of electronic
interferometers~\cite{Ji:2003-1,Neder:2007-1,Neder:2007-2} in quantum
Hall edge channels and finally the 
realization of on-demand single-electron
sources~\cite{Feve:2007-1,Dubois:2013-1}.

Importantly, the
introduction of single-electron sources has catalyzed a shift from questions
about the statistics of charge flowing across a quantum
conductor~\cite{Levitov:1996-1,Blanter:2000-1}
to questions about the wavefunctions of elementary excitations
carrying the charge. This naturally led to the transposition of photon
quantum optics concepts introduced in the 60s by Glauber to coherent
quantum nanoelectronics, thus giving birth to the central concepts of 
electronic coherences~\cite{Degio:2011-1,Haack:2012-2,Moskalets:2014-1,Thibierge:2016-1}.

Historically, the scattering theory approach to quantum 
transport~\cite{Landauer:1985-1,Buttiker:1985-1,%
Buttiker:1986-1,Martin:1992-1,Buttiker:1992-1}
had already emphasized
the importance of optics concepts in electronic
transport. 
However, electron quantum optics differs
from photon quantum optics because of the Fermi statistics of electrons
which changes the nature of the reference ``vacuum state" when all
sources are switched off. Understanding and extending quantum optics
concepts in the presence
of such non-trivial vacua was also a motivation for developing
electron quantum optics.
Even more importantly, electrons are charged particles interacting through 
Coulomb interactions. As stressed out by
M.~B\"{u}ttiker and his 
collaborators~\cite{Buttiker:1993-1,Pretre:1996-1,Christen:1996-1,Blanter:2000-1},
this plays a crucial role in
high frequency quantum transport by enforcing charge conservation and
gauge invariance~\cite{Pretre:1996-1,Blanter:2000-1}.
Coulomb interactions also raise the basic
question of the fate of electronic quasi-particles in a
metal, which was the starting point for the
Landau-Fermi liquid theory~\cite{Nozieres-Pines}.

Electron quantum optics thus offers unprecedented possibilities to study these
basic condensed matter physics questions down to the single-electron level.
These possibilities are
illustrated by recent two-particle
interferometry experiments in quantum Hall edge channels which are
reviewed in the present volume~\cite{Marguerite:2016-2}.

Electron quantum optics also establishes a
bridge between quantum coherent nanoelectronics and microwave quantum optics, which
now plays an important role in superconducting nanocircuits used for
quantum information processing and
manipulation~\cite{Eichler:2011-1,Lang:2013-1,Bozyigit:2011-1}.
Microwave quantum optics is also crucial for understanding the
electromagnetic radiation emitted by a quantum conductor, an important
problem rising a growing
interest~\cite{Beenakker:2004-1,Grimsmo:2016-1,Mendes:2015-1,Forgues:2014-1,Forgues:2015-1,Thibaut:2015-1,Virally:2016-1}.

The purpose of this paper is to reconsider electron quantum optics from
a more global perspective by interpreting it in the language of signal
processing. In its broadest acceptance, signal processing is 
an enabling technology 
that aims at processing, transferring and retrieving
information carried in various physical formats called
``signals"~\cite{Moura:2009-1}. Signal
processing involves a huge arsenal of techniques to detect,
filter, represent, transmit, and finally extract information or
recognize patterns within signals.
However, the most famous and historically important examples of signals,
 such as acoustic or electronic signals, 
are classical.
Here, we would like
to emphasize that all electron optics experiments realized so
far~\cite{Bocquillon:2013-1,Jullien:2014-1,Freulon:2015-1,Marguerite:2016-1} as
well as proposals for accessing two-electron
coherence~\cite{Thibierge:2016-1} can be interpreted
in the signal processing language as experiments on
 signals which are no longer classical:
namely electron quantum optics coherences. Electronic
interferometers realize analog signal processing operations such as
``linear filtering" or ``overlaps"
on these quantum signals and encode the result into experimentally
accessible quantities such as average current and current correlations.

By emphasizing this point of view, we provide a unified view of
the recent developments of electron quantum optics and, as we will show
by discussing two-particle interferometry, we gain some inspiration
towards envisioning new experimental measurement
schemes for electronic coherences. Although this has not really
been fully exploited yet, it could also suggest new ways of obtaining data by
using suitable sources and data processing for optimizing electronic
coherence reconstruction~\cite{Ferraro:2013-1}. It may also rise 
interest in the
signal processing community towards 
electron quantum optics and lead
to new innovative experimental and theoretical ideas.

In order to develop this point of view, this paper is organized as follows: 
after briefly recalling the experimental and theoretical context 
of electron quantum optics in section \ref{sec:context}, 
section \ref{sec:G1} reviews 
the notion of single-electron coherence and the various ways 
to access this quantity. We will discuss the
signal processing operations performed in various single- and 
two-particle interferometers used to probe single-electron coherence.
Our discussion is complementary to that of the review
\cite{Marguerite:2016-2} which discusses two-particle interferometry
experiments in quantum Hall edge channels and the
underlying theory. Here, the same concepts are reviewed with a strong
emphasis on the quantum
signal processing point of view.

We then turn to two-electron coherence in section \ref{sec:G2}, and
present its definition and its main properties. We introduce 
its various representations and discuss its
non-classical features. We then present its relation to quantum noise of
the electrical current, showing the electron quantum optics version of
Einstein's relation between particle number fluctuations and
wavefunctions.
We show that a whole class of experiments can be
interpreted as linear filtering converting the intrinsic second order
coherence emitted by a source into current correlations. 

In the last
section, we will connect electronic coherences to quantum information theoretical
quantities. For this purpose, we will explain how to derive effective 
qubit density matrices from a
set of orthogonal single-particle states. In particular, 
we will
illustrate this by discussing electron/hole entanglement in the
many-body state generated by a mesoscopic capacitor. We will 
finally sketch
how ideas from signal processing lead to a suitable definition of
these density matrices for periodically driven systems. 

\section{The context of electron quantum optics}
\label{sec:context}

Let us briefly review the main steps that have
lead to the development of 
electron
quantum optics.

\subsection{Experiments}
\label{sec:context:experiments}

On the experimental side, the integer quantum Hall effect in high
mobility 2-dimensional electron gas (2DEG) in AsGa/AsGaAl heterostructures provides the analogous of
optical fibers through chiral propagation of charge carriers within the
so-called %#KLH
edge channels. Progresses in nanofabrication not only enabled the fabrication
of the quantum point contact (QPC), which plays the role of an ideal electronic beam
splitter~\cite{Wharam:1988-1,vanWees:1988-1,vanHouten:1992-1}, but 
have also enabled its
embedding into complicated geometries 
such as the electronic Mach-Zehnder
interferometer (MZI)~\cite{Ji:2003-1,Roulleau:2007-2,Roulleau:2008-1} and 
the Samuelsson-B\"{u}ttiker interferometer~\cite{Samuelsson:2004-1,Neder:2007-2}. 
These pioneering experiments
showed that, at low temperatures, electronic
coherence can be maintained over distances comparable to the size of
these circuits (from few
to \SI{20}{\micro\meter})~\cite{Roulleau:2008-2}.
They have also opened the way to the demonstration and
study of more complex electronic circuits in which elements
such as QPC, quantum dots and single-electron sources 
could be placed like optical components
on an optical table. As mentioned in the Introduction, the demonstration
of the mesoscopic capacitor as an on-demand single-electron source
\cite{Feve:2007-1} marked the beginning of electron quantum optics. Now,
other single-electron sources have been demonstrated in
AsGa/AsGaAl, from turnstiles
\cite{Blumenthal:2007-1,Fletcher:2013-1}, mainly motivated by metrology, 
to the Leviton source
\cite{Dubois:2013-1} which is 
reviewed in this volume with great details \cite{Glattli:2016-1}.

Around the same time, progresses in microwave technology
and measurement
techniques have lead to the exploration of high-frequency
transport~\cite{Gabelli:2006-1,Gabelli:2007-1}, thereby
confirming the predictions by B\"uttiker and his
collaborators~\cite{Buttiker:1993-2,Pretre:1996-1}.
From a broader
perspective, these developments have
opened the way to in-depth experimental investigations of high-frequency
quantum coherent nanoelectronics.

\subsection{Theory}
\label{sec:context:theory}

The electron
quantum optics formalism was then developped taking advantage of the
chirality of electronic transport in quantum Hall systems: in such
systems, the optical analogy is exploited at its best by decomposing the
quantum conductor, or more generally the electronic circuit, into simple
building blocks such as quantum point contacts
or energy filters so that
an incoming electronic flow is transformed into an outgoing one. Then,
within the measurement stage, the average 
current or the current noise is measured either at zero
or at a given high frequency. The resulting formalism bears a
close similarity with the input-output formalism of photon
quantum optics~\cite{Gardiner:1985-1}. 
In particular, it assumes
that electronic coherences are probed within a region where interactions
can be neglected. In such a region, electrons propagate freely at a
Fermi velocity which will be denoted by $v_F$ throughout the present
paper. When electronic
coherences are probed at a given position, often corresponding to the
position of a detector, they depend on time variables. 

Such a description contains an
assumption on the screening of Coulomb interactions. For example, we assume that
screening at a quantum point contact is good enough to neglect any
capacitive coupling between the incoming and outgoing edge channels. As
of today, there has been no experimental evidence contradicting this
assumption. 
Within this framework, interaction effects have been studied
extensively, starting with MZI. 
We will very brielfy discuss some of these works in section
\ref{sec:G1:MZI} mentioning that interaction effects can lead to a breaking
of the paradigm of 
quantum conductors as linear electron quantum optics components.  We
also refer the
reader interested by interaction induced decoherence effects in Hong Ou Mandel
(HOM) experiments to \cite{Marguerite:2016-2} as well as
\cite{Marguerite:2016-1}. 

\subsection{New systems}
\label{sec:context:new-systems}

The rapid development of electron quantum optics has also catalyzed a stream
of works whose purpose is to extend its application range to new
physical systems.

A first line of research deals with extending electron quantum optics to
situations in which interactions potentially lead to a drastic change of
the ground state such as in superconductivity. 
This has led to study of electron quantum optics with Bogoliubov
quasi-particles
which is reviewed in this volume
\cite{Ferraro:2016-1}.  Another question is the generalization
of electron quantum optics to fractional quantum Hall (FQH) edge
channels. Proposals have been made for single quasi-particle and
single electron emitted in these systems \cite{Ferraro:2015-1} and the
HOM experiment with Lorentzian pulses has been considered \cite{Rech:2016-1}.
However,
a full generalization of electron quantum optics in the FQH regime is
still missing, the main problem being the absence of any ideal
quasi-particle beam splitter. Nevertheless, perturbative approaches may
prove to be useful for experiments whenever the crossover energy scale
associated with a constriction \cite{Fendley:1995-1} 
is well below the experimentally relevant
energy scales.

Another line of research focuses on manipulating the spin degree of freedom at
the single-electron level. Quantum spintronics has risen a strong interest in
the mesoscopic
physics community because of the importance of coherent spin transport
and manipulation for quantum
information processing.
Although the $\nu=2$ edge channel system had already been envisioned for
quantum spintronics \cite{Karmakar:2011-1}, 2D
topological insulators (TI), such as quantum spin Hall systems \cite{Hasan:2010-1},
are now considered as
a potentially important class of systems for electron quantum optics.
In these materials, such as
CdTe/HgTe and InAs/GaSb
quantum wells at zero magnetic field, edge channels are topologically protected from
backscattering. They come as counterpropagating pairs with opposite spin
polarization (spin-momentum locking).
The mesoscopic capacitor built from a 2D TI has been
proposed as an on-demand single Kramer pair source emitting two single-electron
excitations with opposite spins
\cite{Hofer:2013-1,Inhofer:2013-1}. The HOM experiment has been 
discussed with
such sources \cite{Ferraro:2014-2}. 
A variant of this source based on a driven antidot has also recently been proposed
\cite{Dolcetto:2016-1} as well as
a different system relying on two quantum dots coupled to a 2D TI via
tunneling barriers \cite{Xing:2014-1}.
Coulomb interactions are expected to play an important role in these
systems due to the spatial superposition of two counter-propagating edge
channels. Comparing to the $\nu=2$ quantum Hall edge channel system,
interactions among counterpropagating edge systems are expected to
induce new effects ranging from resonance effects
\cite{Sukhorukov:2007-2} to
fractionalization \cite{Calzona:2016-1}. 

Although these developments go beyond the scope of the present paper, it is important
to keep in mind that the basic concepts of electron quantum optics can
be extended to quantum spintronics in a relatively simple way.

\section{Single-electron coherence}
\label{sec:G1}

Let us now review the concept of single-electron
 coherence and discuss how it can be accessed through single-particle
interferometry and two-particle interferometry. Our main
message is that, under certain circumstances, 
the first type of experiments correspond to linear
filtering in the signal processing language whereas the 
HOM experiment
performs an analog computation of the overlap of two single-electron
coherences.

\subsection{Definition and representations}
\label{sec:G1:definitions}

Single-electron coherence~\cite{Degio:2011-1} is defined by analogy with first-order
coherence in photon quantum optics~\cite{Glauber:1963-1}:
\begin{equation}
\label{eq:G1:difinition}
\mathcal{G}_{\rho}^{(e)}(1|1')=\mathrm{Tr}(\psi(1)\,\rho\,\psi^\dagger(1'))
\end{equation}
where $\rho$ denotes the reduced density operator for the electronic
fluid, $\psi$ and $\psi^\dagger$ denote the fermionic destruction and
creation field operators and 
$1=(\alpha,x,t)$ and $1'=(\alpha',x',t')$ denote edge channel indices
and space time coordinates. For spin
polarized edge channels, $\alpha$ and $\alpha'$ correspond to spin indices:
$\alpha=\alpha'$ then encodes spin populations whereas
$\alpha=-\alpha'$ corresponds to spin coherences, thus showing that 
the formalism of
electron quantum optics can be easily extended to account for the
spin.

In the following, to simplify notations, channel/spin indices will
be dropped out when only one edge channel is involved.
single-electron coherence contains all
the information on single-electron wavefunctions present in the
system. For example, let us consider the $N$ electron state
$|\Psi_N\rangle=
\prod_{k=1}^N\psi^\dagger[\varphi_k]|\emptyset\rangle $ 
where
the creation operator for an electron in the single-particle
state $\varphi_k$ is defined by:
\begin{equation}
\psi^\dagger[\varphi_k]=v_F\int_{\mathbb{R}} \varphi_k(t)\psi^\dagger(t)\,\md t\,.
\end{equation}
This state is obtained by 
filling
mutually orthogonal single-particle states $(\varphi_k)_{k\in
\{ 1,\dots,N \}}$ on top of the electronic vacuum $|\emptyset\rangle$. Then,
a straightforward application
of Wick's theorem shows that, in the space domain at the initial time: 
\begin{equation}
\mathcal{G}^{(e)}_{|\Psi_N\rangle}(x|y)=\sum_{k=1}^N\varphi_k(x)\,\varphi_k(y)^*\,.
\end{equation}
In a metallic conductor, at zero temperature and with all electronic
sources switched off,
the reference state is a Fermi sea vacuum of given chemical potential 
$\mu(x)$. Therefore, contrary to the case of photon quantum optics, the
single-particle coherence does not vanish when sources are switched off 
but reduces to the Fermi sea
contribution 
$\mathcal{G}^{(e)}_{F_{\mu(x)}}(t|t')=\langle
F_{\mu(x)}|\psi^\dagger(x,t')\psi(x,t)|F_{\mu(x)}\rangle$. 
On the contrary, the inter-channel single-electron coherence vanishes when all
electronic sources are switched off:
$\mathcal{G}^{(e)}_{F}(\alpha,t|\alpha',t')=0$ for $\alpha\neq\alpha'$.

In electron quantum optics, $\mathcal{G}_\rho^{(e)}$ is considered at a
given position $x$ within the electronic circuit, thus leading to a
two-time function
$\mathcal{G}_{\rho,x}^{(e)}(t|t')=\mathcal{G}_{\rho}^{(e)}(x,t|x,t')$.
When the sources are switched on, the single-electron coherence is
different from the Fermi sea contribution 
and the excess single-electron coherence is defined by subtracting the
Fermi sea contribution:
$\Delta\mathcal{G}^{(e)}_{\rho,x}(t|t')=
\mathcal{G}_{\rho,x}^{(e)}(t|t')-\mathcal{G}^{(e)}_{F_{\mu(x)}}(t|t')$.

The most convenient representation for single-electron coherence is
a mixed time-frequency representation called the electronic
Wigner distribution function, which captures both
temporal evolution and the nature of excitations~\cite{Ferraro:2013-1}:
\begin{equation}
\label{eq:G1:Wigner}
W^{(e)}_{\rho,x}(t,\omega)=\int_{-\infty}^{+\infty}
v_F\,\mathcal{G}^{(e)}_{\rho,x}
\left(t+\frac{\tau}{2}\Big\vert t-\frac{\tau}{2}\right)\,\me^{\mi\omega\tau}\md\tau
\end{equation}
where $v_F$ denotes the Fermi velocity at position $x$.
The electronic Wigner distribution function is real.
Its marginal distributions give access to the time-dependent average
electric current and to the electronic distribution function
$f_e(\omega)$ at position $x$:
\begin{subequations}
\label{eq:G1:marginals}
\begin{align}
\label{eq:G1:marginals:occupation}
f_e(\omega)&=\overline{W^{(e)}_{\rho,x}(t,\omega)}^t\\
\label{eq:G1:marginals:current}
\langle i(x,t)\rangle_\rho &= -e\int_{-\infty}^{+\infty} \Delta
W^{(e)}_{\rho,x}(t,\omega)\,\frac{\md\omega}{2\pi}
\end{align}
\end{subequations}
A classicality criterion for the electronic 
Wigner distribution function has been formulated \cite{Ferraro:2013-1}:
$0\leq W^{(e)}_{\rho,x}(t,\omega)\leq 1$.
When verified, it basically means that $W^{(e)}_{\rho,x}(t,\omega)$ can be
interpreted as a time-dependent electronic distribution
function. This is the case for a
quasi-classically driven Ohmic contact, when $|eV_{\mathrm{ac}}|\gg
hf$ and $k_B T\gg hf$, $f$ being the driving 
frequency. In this case, both thermal
fluctuations and the ac drive are responsible for multiphoton
transitions in the electronic fluid and many electron/hole pairs are
generated.
On the contrary, for a single-electron excitation, quantum
delocalization is responsible for negativities as seen on the example
of the MZI~\cite{Ferraro:2013-1}. Accessing single-electron coherence would
thus enable to discuss the non classicality of the electronic fluid at 
single-particle level.

\subsection{Single-particle interferometry as linear filtering}
\label{sec:G1:MZI}

\paragraph{Introduction}

In classical signal processing, 
linear filters transform time-dependent input signals into output signals
under the constraint of linearity. Well known examples include linear
circuit elements in classical electronics and linear optics elements 
such as lenses, beam splitters
and other various optical devices. These components act as linear filters on the electromagnetic
field classical coherence introduced in the 30s~\cite{Zernicke:1938-1}. 
This statement also extends to quantum optics by considering 
quantum optical coherences introduced by
Glauber~\cite{Glauber:1963-1}. 

In this section, we show that the same statement
is true in electron quantum optics provided we use quantum conductors in
which interaction effects can be neglected. As an example, we will see
how the ideal Mach-Zenhder interferometer~\cite{Haack:2011-1} or the measurement of the
electronic distribution function using a quantum dot as an energy
filter~\cite{Altimiras:2010-1} realize linear filtering of the excess single-electron coherence 
$\Delta\mathcal{G}_{\rho,x}^{(e)}(t|t')$, which should therefore be seen as a
``quantum signal" depending on two times. 

We will then discuss how Coulomb interactions partly -- but not totally
-- invalidate this
statement. In particular, we will explain why measuring finite-frequency
currents enables to keep track of interaction effects and
to recover information on single-electron coherence.

\paragraph{Mach-Zehnder interferometry}

\begin{figure}
%\missingfigure{A simple figure for the MZI.}
% Figure MZI Etienne
\begin{center}
\begin{tikzpicture}[ scale=0.35,
   decoration={
    markings,% switch on markings
    mark=at position .15 with {\arrow{stealth}},
	mark=at position .5 with {\arrow{stealth}}}
	]

\draw [ultra thick] (-1,0) node [left] {$A$} -- (1,0);
\draw [ultra thick] (7,0) -- (9,0) node [right] {$B$};

\draw [myred, postaction={decorate}] (-3,3) node [above left] {1in} 
	-- (0,0)
	.. controls (2,4) and (6,4) ..
	(8,0) -- ++(3,3) node [above right] {1out};
\draw [myred] (4,3.8) node {1\textsc{mzi}};
\draw (4,2.2) node {$\tau_1$};
\draw [myred, postaction={decorate}] (-3,-3) node [below left] {2in} 
	-- (0,0)
	.. controls (2,-3) and (6,-3) ..
	(8,0) -- ++(3,-3) node [below right] {2out}; 
\draw [myred] (4,-3.2) node {2\textsc{mzi}};
\draw (4,-1.5) node {$\tau_2$};

\draw (4,0) node {$\Phi_B$} ;

\end{tikzpicture}
\end{center}
\caption{\label{fig:MZI} Schematic view of the Mach-Zehnder
interferometer: the incoming channels are partitionned at the electronic
beam splitter $A$ and then recombined by the beam splitter $B$. Here
$\tau_1$ and $\tau_2$ denote the times of flight across the two
branches of the MZI and $\Phi_B$ the magnetic flux enclosed by the
interferometer.}
\end{figure}
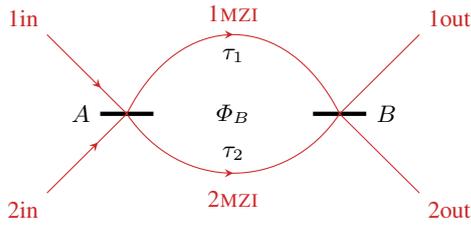

An ideal electronic Mach-Zehnder interferometer, such as the one depicted on
Fig.~\ref{fig:MZI}, is characterized by the time of flights $\tau_{1,2}$
along its two arms and the magnetic flux threading it $\Phi_B=\phi_B\times
(h/e)$. When an electronic source $S$ is placed on the incoming edge channel $1$, 
the time-dependent outgoing electric current in channel $1$ is directly
proportional to the excess electronic coherence of the
source~\cite{Haack:2011-1,Ferraro:2013-1}:
\begin{subequations}
\begin{align}
\label{eq:MZI:classical}
&\langle i_{1\mathrm{out}}(t)\rangle = \sum_{j=1,2}\mathcal{M}_{j,j}
\langle i_S(t-\tau_j)\rangle\\
\label{eq:MZI:quantum}
&-2e|\mathcal{M}_{1,2}|\int_{\mathbb{R}}\cos{(\omega \tau_{12}+\phi)}\,\Delta
W_S^{(e)}(t-\bar{\tau},\omega)\,\frac{\md\omega}{2\pi}\,
\end{align}
\end{subequations}
where $\mathcal{M}_{i,j}$ denotes the product
$\mathcal{A}_i\mathcal{A}_j^*$, $\mathcal{A}_j$ 
being the transmission amplitude of the beam splitters
along path $j$ of the interferometer. We have introduced
$\tau_{12}=\tau_1-\tau_2$, $\bar{\tau}=(\tau_1+\tau_2)/2$
and $\phi=\mathrm{arg}(\mathcal{M}_{1,2}+2\pi\phi_B)$ which is the phase
associated with both beam splitters and the magnetic flux. 
The first line (Eq.~\eqref{eq:MZI:classical}) does not depend on the magnetic flux and
therefore corresponds to classical
propagation within each of the two arms of the MZI, whereas the second line
(Eq.~\eqref{eq:MZI:quantum}) corresponds to quantum interferences between
propagations within both arms.
Because the average electric
current is also proportional to the excess single-electron coherence of
the source, the outgoing average current is obtained from the
excess incoming coherence $\Delta\mathcal{G}^{(e)}_S$ in channel
$\mathrm{1in}$ 
by a linear filter which we write symbolically
$\langle i_\text{out,dc}\rangle =
\mathcal{L}_{\text{MZI}}[\Delta\mathcal{G}^{(e)}_S]$.
Measurements of the $\Phi_B$ dependent part of the average dc
current for various $\tau_1-\tau_2$ could then be used to reconstruct single-electron
coherence~\cite{Haack:2011-1}.

The key ingredient in this derivation is the absence of electronic
interactions. 
Whenever one replaces the Mach-Zehnder interferometer by
an ideal ballistic quantum conductor in which interactions are neglected, the
outgoing current in the measurement lead would also be 
proportional to $\Delta \mathcal{G}_S^{(e)}$. 
Denoting by
$\mathcal{S}(t_f,t_i)$ the scattering amplitude for an electron 
arriving into the conductor at time $t_i$ and going out towards the
measurement lead at time $t_f$, then 
the outgoing average time-dependent current is given by
\begin{equation}
\langle i_{\mathrm{out}}(t)\rangle = 
\int_{\mathbb{R}^2}
\mathcal{S}(t,t_+)\,\mathcal{S}^*(t,t_-)\Delta
\mathcal{G}_{S}^{(e)}(t_+,t_-)\,\md t_+\md t_-\,
\end{equation}
which describes a linear filtering of the incoming single-electron
coherence $\Delta\mathcal{G}^{(e)}_S$
associated with time-dependent scattering. In particular, this expression is
valid within 
the framework of Floquet scattering
theory~\cite{Moskalets:book}.

\paragraph{On the role of interactions}

However, as we shall now discuss, because of Coulomb interactions,
the situation in electron quantum optics is subtler than in
photon quantum optics. It is therefore important to clarify
in which
situations the linear filtering paradigm is valid and when
it is not valid.

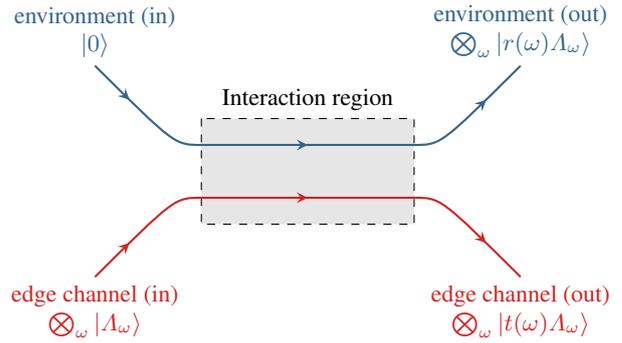
\begin{figure}
%\missingfigure{A figure for the usual interaction stuff: incoming state,
%environmental channels, interaction region, outgoing state}
\begin{center}
% Figure région d'interactions
\begin{tikzpicture}[>=stealth, scale=0.7,
	decoration={
    markings,% switch on markings
    mark=at position .1 with {\arrow{stealth}},
	mark=at position .5 with {\arrow{stealth}},
	mark=at position .9 with {\arrow{stealth}},
	}]

\filldraw[fill=gray!20, dashed] (-2,1) rectangle (2,-1);

\draw [myblue, thick, postaction=decorate] (-4,2) .. controls  (-2.5,0.5) .. (-2,.5) -- (2,0.5)  .. controls  (2.5,0.5) .. (4,2);
\draw [myred, thick, postaction=decorate] (-4,-2) .. controls  (-2.5,-0.5) .. (-2,-.5) -- (2,-0.5)  .. controls  (2.5,-0.5) .. (4,-2);

\draw (0,1) node [above, align=center] {Interaction region};
\draw [myblue] (-4,2) node [above, align=center] {environment (in)\\  $\ket{0}$};
\draw [myred] (-4,-2) node [below, align=center] {edge channel (in)\\  $ \bigotimes_\omega \ket{\Lambda_\omega}$};
\draw [myblue] (4,2) node [above, align=center] {environment (out)\\
$\bigotimes_\omega  \ket{r(\omega) \Lambda_\omega}$};
\draw [myred] (4,-2) node [below, align=center] {edge channel (out)\\
$\bigotimes_\omega \ket{t(\omega) \Lambda_\omega}$};
\end{tikzpicture}
\end{center}
\caption{\label{fig:interactions} Input/output treatment of a 
finite-length
interaction region (light grey box) in the edge-magnetoplasmon scattering
framework: an edge channel (in red) is capacitively coupled to another
conductor in the linear response regime, 
described by an environmental channel (in blue). At zero
temperature, the incoming environmental modes are in their vacuum state.
When a coherent edge-magnetoplasmon state $\ket{\Lambda} =
\otimes_\omega \ket{\Lambda_\omega}$ is sent in the incoming channel,
the outgoing state is
$\otimes_{\omega}\left(\ket{t(\omega)\Lambda_\omega}\otimes
\ket{r(\omega)\Lambda_\omega}\right)$ where $t(\omega)$
denotes the edge-magnetoplasmon transmission 
amplitudes across the interaction region and $r(\omega)$ the amplitude
to be scattered into enviromental modes. 
}
\end{figure}

To begin with,
let us consider a simple situation in
which the source is placed at the input of a finite-length interaction region,
as depicted on
Fig.~\ref{fig:interactions}. 
For simplicity, we assume that this
interaction region can be described within the edge-magnetoplasmon
scattering formalism~\cite{Safi:1999-1,Degio:2009-1}: the incoming
edge-magnetoplasmon mode at a given frequency is scattered elastically
between
the corresponding outgoing mode and environmental modes associated with 
other electrical degrees of freedom present in the problem. The
environment can involve
edge channels or any neighboring linear circuit modeled as
a transmission line with a frequency-dependent
impedance~\cite{Yurke:1984-1}. Such a
modelization is valid as long as all the components of the system are in
the linear response regime.
This formalism was used to
compute the effect of Coulomb interactions on coherent current pulses
\cite{Grenier:2013-1} and 
single-electron 
excitations~\cite{Degio:2009-1,Ferraro:2014-1}.
Equivalently, because they induce inelastic processes, Coulomb
interactions turn a quantum conductor into a non-linear component from
the electron quantum optics point of view. 
Consequently, in general, the excess outgoing single
electron coherence is not a linear filtering of the incoming one!

A criterion for
the validity of the electronic scattering theory approach to quantum
transport at finite frequencies is that
the frequency dependence of the electronic scattering matrix
of a quantum conductor can be neglected~\cite{Blanter:2000-1}. Single- to 
few-electron excitations emitted by electron quantum optics
sources such as the mesoscopic capacitor~\cite{Feve:2007-1} or the
Leviton source~\cite{Dubois:2013-1} as well as 
periodic electric currents generated using an advanced waveform
generator usually define a frequency scale 
in the range of one to few tens of \SI{}{\giga\hertz}. On the other hand, an extended
conductor such as a MZI has  a scattering matrix varying over frequency
scales of the order of the inverse of the time of flight of the
conductor. For a $\SI{10}{\micro\meter}$ interferometer, it is of the order
of $\SI{10}{\giga\hertz}$.
This is why extended quantum
conductors such as MZIs 
fail to satisfy this
criterion. The important stream of theoretical work on interaction-induced decoherence~\cite{Chalker:2007-1,Neder:2007-4,Levkivskyi:2008-1,Neuenhahn:2008-1,Kovrizhin:2009-1} 
in MZI interferometers illustrate
the full complexity of understanding interaction effects in such 
extended quantum conductors.
More recent 
works~\cite{Tewari:2016-1,Slobodeniuk:2016-1} dealing with the
propagation of
individualized energy-resolved single-electron excitations in a MZI are
directly relevant for electron quantum optics but also show that
this problem is not yet fully understood even at the single-electron
level.

By contrast,
Coulomb interaction effects can be 
neglected over a much broader frequency range in the QPC
which is an almost point-like electronic beam-splitter.
As we shall see in the next section, this plays a very important role for the HOM
and HBT experiments.

Last but not least, 
the average finite-frequency currents are relatively robust to
the effect of Coulomb interactions: whenever all electrical components
respond linearily to the incoming excitation, the edge-magnetoplasmon
scattering matrix can be used to reconstruct the incoming average finite
frequency currents from the outgoing ones. 
Measurements of finite-frequency average currents have been successfully
developped in the 1-11~\SI{}{\giga\hertz} range to perform 
the first direct measurement of edge-magnetoplasmon scattering
amplitudes~\cite{Bocquillon:2013-2}.

\subsection{Two-particle interferometry as overlap}
\label{sec:G1:HOM}

\paragraph{Motivation}
\label{sec:G1:HOM:motivation}

Although amplitude interferometry relies on the measurement of average
currents, it does not seem well suited to perform single-electron
tomography. First of all, as in optics, it would require a perfect
control on electronic optical paths down to the 
Fermi wavelength. 
More importantly, as discussed in the previous section, Coulomb
interactions prevent
reconstructing the incoming single-electron coherence from the
experimental signals. 

This situation is very similar to what happened in astronomy in the 30s
and 40s: 
attempts at directly measuring the diameter of normal stars using
amplitude interferometry were plagued by atmospheric turbulences and by
the technological challenge of building a large optical interferometer. 
However, a way to circumvent this bottleneck was
found by Hanbury Brown and Twiss (HBT) in the 50s~\cite{Hanbury:1956-1}: their idea was 
to measure intensity correlations~\cite{Hanbury:1956-2} since these
contain interferences between waves emitted by pairs of atoms on the star.
In quantum optics, the HBT effect
ultimately relies on
two-photon interferences~\cite{Fano:1961-1}.
In the 80s, the Hong Ou
Mandel (HOM) experiment~\cite{Hong:1987-1} also demonstrated two-particle interferences 
for identical particles (photons). Since then, 
two-particle interference
effects have been observed in many different contexts, from stellar
interferometry to nuclear and particle physics~\cite{Baym:1998-1} and more
recently with bosonic as well as fermionic cold
atoms~\cite{Jeltes:2007-1}. Recent experiments demonstrate a higher
degree of control by using independent single-photon~\cite{Beugnon:2006-1} and
single atom sources~\cite{Lopes:2015-1}. 

In this section, we review how the HOM experiment can
be used to measure the overlap of the excess single-particle coherences
arriving at a beam splitter. 
Remarkably, this result is true not only for electrons but for any
fermionic or bosonic excitation. In photon quantum optics, it forms the
basis of homodyne tomography~\cite{Smithey:1993-1,Lvovsky:2009-1} 
recently used to characterize few-photon states in the optical
domain~\cite{Ourjoumtsev:2006-1}. In the microwave domain, the 
HOM scheme has been used to access photon quantum optical
correlations 
from electrical
measurements~\cite{Bozyigit:2011-1,Lang:2013-1} and forms the basis of
a tomography scheme for itinerant microwave
photons~\cite{daSilva:2010-1,Eichler:2011-1}.

\paragraph{Theoretical analysis}
\label{sec:G1:HOM:theory}

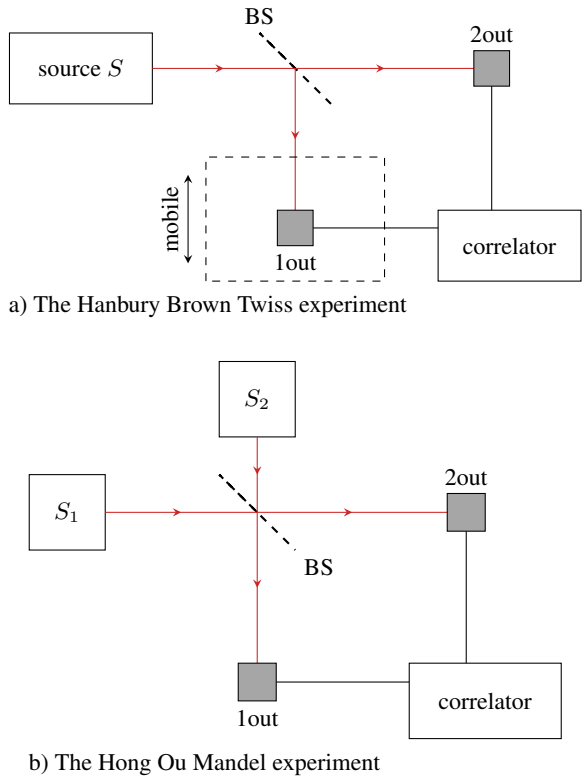
\begin{figure}
\begin{center}
% Principle for HBT experiment
\subfloat[The Hanbury Brown Twiss experiment]{%
\begin{tikzpicture}[scale=.47, >=stealth,
	decoration={
    markings,% switch on markings
    mark=at position .5 with {\arrow{stealth}}}]

\draw (-2,1) rectangle ++ (4,-2);
\draw (0,0) node {source $S$};

\draw [thick, dashed] (6,0) -- ++(-1,1) node [above, align=center] {BS} -- ++(2,-2);

\draw [myred, postaction={decorate}] (2,0) -- (6,0);
\draw [myred, postaction={decorate}] (6,0) -- ++(5,0);
\draw [myred, postaction={decorate}] (6,0) -- ++(0,-4);

\filldraw [fill=gray!70] (5.5,-4) rectangle ++(1,-1);
\draw (6,-5) node [below] {1out};
\filldraw [fill=gray!70] (11,0.5) rectangle ++(1,-1);
\draw (11.5,0.5) node [above] {2out};

\draw (11.5,-.5) -- ++(0,-3.5);
\draw (6.5,-4.5) -- ++(3.5,0);

\draw (10,-4) rectangle ++ (4,-2);
\draw (12,-5) node {correlator};

\draw [dashed] (3.5,-6) rectangle ++(5,3.5);
\draw [<->] (3,-5.5) -- ++(0,2.5);
\draw (2.5,-4.25) node [rotate=90] {mobile};

\end{tikzpicture}
}\\[5mm]
% Schema pour HOM
\subfloat[The Hong Ou Mandel experiment]{%
\begin{tikzpicture}[scale=.5, >=stealth,
	decoration={
    markings,% switch on markings
    mark=at position .5 with {\arrow{stealth}}}]

\draw (0,1) rectangle ++ (2,-2);
\draw (1,0) node {$S_1$};

\draw (5,4) rectangle ++ (2,-2);
\draw (6,3) node {$S_2$};

\draw [thick, dashed] (6,0) -- ++(-1,1) -- ++(2,-2) node [below right] {BS};

\draw [myred, postaction={decorate}] (2,0) -- (6,0);
\draw [myred, postaction={decorate}] (6,2) -- (6,0);
\draw [myred, postaction={decorate}] (6,0) -- ++(5,0);
\draw [myred, postaction={decorate}] (6,0) -- ++(0,-4);

\filldraw [fill=gray!70] (5.5,-4) rectangle ++(1,-1);
\draw (6,-5) node [below] {1out};
\filldraw [fill=gray!70] (11,0.5) rectangle ++(1,-1);
\draw (11.5,0.5) node [above] {2out};

\draw (11.5,-.5) -- ++(0,-3.5);
\draw (6.5,-4.5) -- ++(3.5,0);

\draw (10,-4) rectangle ++ (4,-2);
\draw (12,-5) node {correlator};

\end{tikzpicture}
}
\end{center}
\caption{\label{fig:HBT-HOM:principle} Principle of the HBT and HOM
experiments: in the optical HBT experiment (a), excitations emitted by a
single source $S$ are partitioned
at the beamsplitter $\mathrm{BS}$ whereas in the HOM experiment (b), excitations
emitted by two sources $S_1$ and $S_2$ are sent onto $\mathrm{BS}$. In
optics, one performs a time-resolved detection of photons. In
the electronic case, the beamsplitter is a QPC, and one measures current
correlations between $1\mathrm{out}$ and $2\mathrm{out}$ 
or current noise in the $1\mathrm{out}$ channel. In the case of the HBT
experiment, vacuum is replaced by the reference Fermi sea.}
\end{figure}

In electron quantum optics, the HBT and HOM experiments are demonstrated
by sending electronic excitations generated by one or two electronic
sources on an ideal electronic beam splitter, as depicted on Fig.
\ref{fig:HBT-HOM:principle}. 

In order to make a precise analogy with
photon quantum optics, keeping in mind that in electron quantum optics,
the vacuum is the reference Fermi sea and not a true vacuum, we 
consider that the electronic analogue of the table-top HBT experiment
(Fig.~\ref{fig:HBT-HOM:principle}a)~\cite{Hanbury:1956-2}
is realized when one of the incoming channels is fed with the 
reference Fermi sea ($S_1$ or $S_2$ being off). 
By the same analogy, the electronic HOM experiment
(Fig.~\ref{fig:HBT-HOM:principle}b) corresponds to
situations with both electronic sources in the incoming channels
switched on.
Finally, whereas in photon quantum optics, the arrival of individual photons can
be recorded, in electronics, the quantities of interest are the current
correlations at zero frequency in the two outgoing branches. 

Let us focus on the outgoing current noise in channel~$1$. A first
important point is that the low-frequency current noise does not
depend on the distance to the QPC: interaction effects lead to 
edge-magnetoplasmon 
scattering among the various outgoing edge channels close to the one
considered but the total power remains
the same. This is why HBT/HOM interferometry is immune to
interaction effects in the measurement stage (beyond the QPC). In the same way, the
intensity correlations measured in an optical stellar HBT 
interferometer are not blurred by
atmospheric turbulences.

Consequently, what we need is the excess low frequency current noise just after the QPC
when both
sources $S_1$ and $S_2$ are switched on. It
is the sum of three contributions~\cite{Ferraro:2013-1}:
\begin{align}
\Delta S_{11}^{(S_1\& S_2)}=\Delta S_{11}^{(S_1)} + \Delta
S_{11}^{(S_2)} + \Delta S_{11}^{(\mathrm{HOM})}
\end{align}
where $\Delta S_{11}^{(S_1)}$ and $\Delta S_{11}^{(S_2)}$
are the excess current noise when only the source $S_1$ (resp. $S_2$) is
switched on. These contributions correspond to the excess noise in HBT experiments
performed on each of the sources. The last term
\begin{align}
\Delta & S_{11}^{(\mathrm{HOM})}= \notag \\
&-2e^2RT\int_{\mathbb{R}^2} \Delta
W^{(e)}_{1\mathrm{in}}(t,\omega)
\Delta W^{(e)}_{2\mathrm{in}}(t,\omega)\,\frac{\md t\,\md\omega}{2\pi}
\label{eq:HOM-overlap}
\end{align}
is the overlap of the excess single-electron coherences arriving
at the QPC~\cite{Ferraro:2013-1} ($R$ and $T$ denoting the reflection and
transmission probabilities of the QPC). 
Eq. \eqref{eq:HOM-overlap} encodes the effect of two-particle interferences
between the excitations emitted by these sources. Note that the time delay of
the two sources can be controlled and therefore a single experimental run gives
access to the time-shifted overlaps of the excess electronic Wigner functions
of the two sources. Finally, the minus sign comes from the fermionic
statistics of electrons.

The two first terms also involve two-particle quantum
interferences. Because they contain information on 
coherences of each of the sources, they will be
discussed in section \ref{sec:G2}. Our point here is to emphasize that
the electronic HOM experiment automatically encodes into the
experimental signal  what the signal
processing community would call the sliding inner
product of the quantum signals formed by the incoming excess  
single-electron coherences in channels $1$ and $2$. This is why the HOM
experiment is so important: it
can be used to test for unknown excess
electronic Wigner functions by looking at their overlaps with themselves
or with the ones generated by controlled and calibrated sources. 
This idea has been expanded to describe a generic tomography protocol
for reconstructing an unknown excess single-electron coherence from its
overlaps with coherences generated by suitable ac + dc
drives~\cite{Degio:2010-4}. We refer the reader to
\cite{Marguerite:2016-2} in the same volume for a detailed description 
of this protocol.

\paragraph{Experimental demonstration}
\label{sec:G1:HOM:experiments}

The electronic HBT experiment has been demonstrated in the
late 90s using dc sources~\cite{Oliver:1999-1,Henny:1999-1} and more
recently 
using single-electron
sources~\cite{Bocquillon:2012-1} which were then used to perform the
electronic HOM experiment~\cite{Bocquillon:2013-1}. These
experiments have paved the way to measurements and studies of 
electron decoherence down to the single-electron level through HOM interferometry.

The idea of the tomography protocol has been
recently demonstrated by D.C.~Glattli's
group~\cite{Jullien:2014-1}: in this
experiment, a stream of Lorentzian pulses is sent onto a beam splitter
whose other incoming
channel is fed with a small ac drive on top of a dc bias. Measurement of
the low-frequency noise then enables 
reconstructing the photo-assisted transition amplitudes which, in this
case, contain all the information on single-electron and higher-order electronic
coherences~\cite{Dubois:2013-1}. This experiment being performed in a
2DEG at zero magnetic field, interaction effects can be neglected and
the experiment leads to the reconstruction of the Leviton single-electron
coherence~\cite{Jullien:2014-1}.

As reviewed in this volume~\cite{Marguerite:2016-2},
the HOM experiment has also recently been used to probe interaction effects
within quantum Hall edge channels. In these experiments, performed at
filling factor $\nu=2$, two single-electron sources are
located at some distance of the QPC and interaction effects are strong
enough to lead to quasi-particle
destruction, as suggested by energy relaxation experiments~\cite{LeSueur:2010-1}. 
First, the HOM effect was used to probe
how interactions lead to fractionalization of classical current
pulses~\cite{Freulon:2015-1} in qualitative agreement
with the neutral/charge edge-magnetoplasmon mode
model~\cite{Levkivskyi:2008-1} which had been already probed through energy relaxation
experiments~\cite{Degio:2010-1} and high-frequency admittance
measurements~\cite{Bocquillon:2013-2}.
But the real strength of HOM experiment comes from its ability to probe
electronic coherence in a time- and energy-resolved way.
It was thus recently used to study quantitatively the
effect of Coulomb interactions on energy-resolved single-electron
excitations (called Landau excitations)~\cite{Marguerite:2016-1}.
The experimental data confirm theoretical
predictions and validate the decoherence scenario based on 
edge-magnetoplasmon scattering~\cite{Wahl:2013-1,Ferraro:2014-1}.

\section{Two-electron coherence}
\label{sec:G2}

Let us now turn to two-electron coherence. After briefly reviewing its
definition and 
properties, introducing the two-electron Wigner distribution function
will enable us to emphasize its non-classical features arising from 
Fermi statistics. We will then turn to two-particle interferometry and
show that, under appropriate hypotheses, these experiments perform 
a linear filtering on 
the intrinsic two-electron coherence, thus generalizing what was
discussed in section \ref{sec:G1:MZI}.

\subsection{Definition and representations}
\label{sec:G2:definitions}

\paragraph{Definition}

Two-electron coherence is defined by direct analogy with Glauber's
second-order photonic coherence~\cite{Moskalets:2014-1}:
\begin{equation}
\label{eq:G2:definition}
\mathcal{G}_\rho^{(2e)}(1,2|1'2')=\mathrm{Tr}(\psi(2)\psi(1)\rho\,
\psi^\dagger(1')\psi^\dagger(2'))\,
\end{equation}
where $1=(\alpha_1,x_1,t_1)$ and $2=(\alpha_2,x_2,t_2)$,
$\alpha_{1,2}$ being channel indices, and similarly for $1'$ and $2'$. 
Its physical meaning can be obtained by computing the two-electron
coherence for the
state
$|\Psi_N\rangle$ defined in section \ref{sec:G1:definitions}. The result
is a sum over all the two-electron wavefunctions
$\varphi_{k,l}(x,y)=\varphi_k(x)\varphi_l(y)-\varphi_k(y)\varphi_l(x)$
($k<l$) that can be extracted from the $N$ single-particle
wavefunctions $(\varphi_k)_{k\in \{ 1,\dots,N \}}$~\cite{Thibierge:2016-1}:
\begin{equation}
\label{eq:G2:wavefunctions}
\mathcal{G}^{(2e)}_{|\Psi_N\rangle}(x_1,x_2|x'_1,x'_2)=
\sum_{k<l}\varphi_{k,l}(x_1,x_2)\,\varphi_{k,l}(x'_1,x'_2)\,.
\end{equation}
Two-electron coherence at a given
position is  a function of four times
$(t_1,t_2;t'_1,t'_2)$. 

\paragraph{Fermionic statistics and two-electron coherence}

Although its definition and Eq.~\eqref{eq:G2:wavefunctions} may suggest
that two-electron coherence
bears similarity with single-electron coherence discussed in the
previous section, it already contains all the counter-intuitive aspects
of quantum indiscernability.
Electrons being fermions, the two-electron coherence satisfies the
following
antisymmetry properties:
\begin{subequations}
\label{eq:G2:antisymmetry}
\begin{align}
\mathcal{G}^{(2e)}_\rho(1,2|1'2')&=-\mathcal{G}_\rho^{(2e)}(2,1|1',2')\\
&=-\mathcal{G}_\rho^{(2e)}(1,2|2',1')
\end{align}
\end{subequations}
Antisymmetry leads to the global symmetry of two-electron coherence
$\mathcal{G}_\rho^{(2e)}(1,2|1',2')=\mathcal{G}_\rho^{(2e)}(2,1|2',1')$ expressing
that the order of detection of electrons does not matter. It also implies that
two-electron coherence vanishes whenever $1=2$ or $1'=2'$: this 
is the Pauli exclusion principle. Together with the
conjugation relation 
\begin{equation}
\mathcal{G}^{(2e)}_\rho(1,2|1',2')^*=\mathcal{G}^{(2e)}_\rho(1',2'|1,2)\,,
\end{equation}
antisymmetry implies that two-electron coherence is defined by its
values for only $1/8$-th of the possible arguments.

The intrinsic two-electron coherence emitted by a source can be defined
from the second-order electronic coherence by subtracting not only the
Fermi sea contribution but also all processes contributing to two-electron
 detection and involving the excess single-electron coherence
of the source. These involve classical contributions as well as quantum
exchange contributions~\cite{Thibierge:2016-1}:
\begin{subequations}
\label{eq:G2:decomposition}
\begin{align}
\mathcal{G}&_\rho^{(2e)}(1,2|1',2') = \notag \\
\label{eq:G2:decomposition:Fermi}
 &\mathcal{G}_F^{(2e)}(1,2|1',2')\\
\label{eq:G2:decomposition:classical}
+& \mathcal{G}^{(e)}_F(1|1')\,\Delta\mathcal{G}_\rho^{(e)}(2|2')+
\mathcal{G}^{(e)}_F(2|2')\,\Delta\mathcal{G}_\rho^{(e)}(1|1')\\
\label{eq:G2:decomposition:exchange}
-&\mathcal{G}^{(e)}_F(1|2')\,\Delta\mathcal{G}_\rho^{(e)}(2|1')
-\mathcal{G}^{(e)}_F(2|1')\,\Delta\mathcal{G}_\rho^{(e)}(1|2')\\
\label{eq:G2:decomposition:intrinsic}
+& \Delta\mathcal{G}_\rho^{(2e)}(1,2|1',2')\,.
\end{align}
\end{subequations}
Eq.~\eqref{eq:G2:decomposition}
should be seen as a definition of the intrinsic two-electron
coherence $\Delta\mathcal{G}^{(2e)}_{\rho}$ 
from the total two-electron coherence, the Fermi sea two-electron
coherence and lower-order electronic coherences.
The second term \eqref{eq:G2:decomposition:classical} is present for classical
particles and represent classical correlations in which the origin of
the two detected particles can be traced back either to the 
Fermi sea or the source.
Such back-tracking is not possible for the exchange terms \eqref{eq:G2:decomposition:exchange} 
whose minus sign
comes from the fermionic
statistics. Note that Eq.~\eqref{eq:G2:decomposition} is fully compatible with
Eq.~\eqref{eq:G2:wavefunctions}. Moreover, for a state obtained by adding a
single-electron or hole excitation to the Fermi sea, the intrinsic 
two-electron coherence vanishes as
expected for a source emitting only one excitation.

\paragraph{The frequency domain representation of two-electron
coherence}

Exactly as in the case of single-electron coherence, the nature of
excitations can be obtained by going to the frequency domain:
\begin{align}
\label{eq:G2:frequency-domain}
\widetilde{\mathcal{G}}^{(2e)}_\rho&(\boldsymbol{\omega}_+|\boldsymbol{\omega}_-)
\notag \\
=&\int_{\mathbb{R}^4}
\mathcal{G}^{(2e)}_\rho(\mathbf{t}_+|\mathbf{t}_-)\,\me^{\mi(\boldsymbol{\omega}_+\cdot\mathbf{t}_+-
\boldsymbol{\omega}_-\cdot\mathbf{t}_-)}\,\md^2\mathbf{t}_+\md^2\mathbf{t}_-\,
\end{align}
where $\mathbf{t}_+=(t_1,t_2)$ and $\mathbf{t}_-=(t'_1,t'_2)$ are
respectively conjugated to $\boldsymbol{\omega}_+=(\omega_1,\omega_2)$ and
$\boldsymbol{\omega}_-=(\omega'_1,\omega'_2)$. Note that antisymmetry
properties \eqref{eq:G2:antisymmetry} are also
true in the frequency domain. 

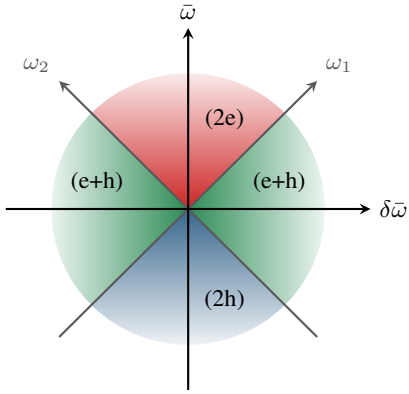
\begin{figure}
\begin{center}
% Figure pour decomposition plan diagonal G2.
\begin{tikzpicture}[scale=1.2]

\begin{scope}[rotate=45]
\shade[top color=white, bottom color=myred] (0,0) -- ++(1.5,0) arc (0:90:1.5)-- cycle;
\shade[top color=myblue, bottom color=white] (0,0) -- ++(-1.5,0) arc (-180:-90:1.5)-- cycle;
\shade[left color=mygreen, right color=white] (0,0) -- ++(1.5,0) arc (0:-90:1.5)-- cycle;
\shade[right color=mygreen, left color=white] (0,0) -- ++(0,1.5) arc (90:180:1.5)-- cycle;

\draw [black!30!gray,thick, ->, >=stealth] (-2,0) -- (2,0) node [above right] {$\omega_1$};
\draw [black!30!gray,thick, ->, >=stealth] (0,-2) -- (0,2) node [above left] {$\omega_2$};
\end{scope}

\draw [thick, ->, >=stealth] (-2,0) -- (2,0) node [right] {$\delta\bar{\omega}$};
\draw [thick, ->, >=stealth] (0,-2) -- (0,2) node [above] {$\bar{\omega}$};

\node at (0.4,1) {(2e)};
\node at (0.4,-1) {(2h)};
\node at (-1,.3) {(e+h)};
\node at (1,.3) {(e+h)};

\end{tikzpicture}

\end{center}
\caption{
\label{fig:G2:excitations} Nature of two-particle excitations:
partitioning the diagonal plane
$(\omega_1,\omega_2)=(\bar{\omega}+\delta\bar{\omega}/2,\bar{\omega}
-\delta\bar{\omega}/2)$ in two sectors associated with pairs of electrons
(2e)
and pairs of holes (2h) and two sectors associated with electron/hole pairs
(e+h).}
\end{figure}

The diagonal of the frequency domain
($\boldsymbol{\omega}_-=\boldsymbol{\omega}_+=(\omega_1,\omega_2)$) 
can be divided into 
quadrants depicted on Fig.~\ref{fig:G2:excitations} that describe the elementary two-particle excitations.
When $\omega_1$ and $\omega_2$ are both positive, we have an electronic
pair whereas when they are both negative, we have a pair of holes. In
the case one is positive and the other negative, we have an electron
hole pair. Note that this classification is compatible with the
permutation $\omega_1\leftrightarrow \omega_2$.

The full frequency domain $(\boldsymbol{\omega}_+,\boldsymbol{\omega}_-)$ 
can then be decomposed into 4D simplexes based on
these quadrants for the diagonal. This will naturally
be compatible with the antisymmetry properties of the two-electron
coherence. Diagonal simplexes are based on
$\boldsymbol{\omega}_+$ and $\boldsymbol{\omega}_-$ that both describe the same type of
excitation. This leads to a two-electron simplex, a two-hole simplex and
two electron/hole pair simplexes respectively containing the contributions of
two-electron, two-hole and electron/hole pair excitations. The
off-diagonal simplexes where $\boldsymbol{\omega}_+$ and $\boldsymbol{\omega}_-$
do not belong to the same quadrant describe coherences between these
four different two-particle excitations.

\subsection{The Wigner representation of two-electron coherence}

\paragraph{Definition}

The Wigner representation of two-electron coherence is defined in the
same way as for single-electron coherence, that is as a Fourier
transform with respect to the time differences $\tau_j=t_j-t'_j$. When
considering a diagonal two-electron coherence in the channel indices
($\alpha_j=\alpha'_j$ for all $j=1,2$), this leads to a real
function 
\begin{align}
\label{eq:G2:Wigner:definition}
W^{(2e)}_{\rho,x}&(t_1,\omega_1;t_2,\omega_2)= \notag \\
&\int_{\mathbb{R}^2}
v_F^2
\mathcal{G}^{(2e)}_{\rho,x}\left(\boldsymbol{t}+\frac{\boldsymbol{\tau}}{2}\Big\vert
\boldsymbol{t}-\frac{\boldsymbol{\tau}}{2}\right)\,\me^{\mi\boldsymbol{\omega}\cdot
\boldsymbol{\tau}}\,\md^2\boldsymbol{\tau}\,.
\end{align}
The Wigner representation of the excess two-electron coherence 
$\Delta
W^{(2e)}_{\rho,x}(t_1,\omega_1;t_2,\omega_2)$ is defined by Eq.
\eqref{eq:G2:Wigner:definition} from 
the excess two-electron coherence.
Whenever Wick's theorem applies, the total two-electron coherence can be
computed in terms of the single-electron one:
\begin{align}
\label{eq:G2:Wick}
\mathcal{G}_{\rho,x}^{(2e)}&(1,2|1',2')=\notag \\
&\mathcal{G}^{(e)}_{\rho,x}(1|1')\,\mathcal{G}^{(e)}_{\rho,x}(2|2')
-\mathcal{G}^{(e)}_{\rho,x}(1|2')\,\mathcal{G}^{(e)}_{\rho,x}(2|1')\,.
\end{align}
and, using Eqs. \eqref{eq:G2:decomposition}, the same equation also
describes the intrinsic two-electron coherence in terms of the
 excess single-electron coherence.
The first term contributes to the two-electron Wigner distribution
through the product
$W^{(e)}_{\rho,x}(t_1,\omega_1)\,W^{(e)}_{\rho,x}(t_2,\omega_2)$ which
corresponds to independent classical particles. The second term comes
from quantum exchange and, as we shall see now, is responsible for
non-classical features of the two-electron Wigner distribution function.

\paragraph{Non classicality of two-electron coherences}

In the case of single-electron coherence, a definition of classicality
was given~\cite{Ferraro:2013-1} based on the idea of interpreting
$W_{\rho,x}^{(e)}(t,\omega)$ as a time-dependent
electronic distribution function compatible with the Pauli exclusion
principle. It is natural to extend this definition to the two-particle
case: $W^{(2e)}_{\rho,x}(t_1,\omega_1;t_2,\omega_2)$ would be called classical
if it takes values between $0$ and $1$. Of course, if we consider the
inter-channel two-electron Wigner distribution associated with the
inter-channel two-electron coherence
$\mathcal{G}^{(2e)}_{\rho,x}(1,t_1;2,t_2|1,t'_1;2,t'_2)$, then when the two
channels are not correlated, we have $W^{(2e)}_{\rho,x}(1,t_1,\omega_1;2,t_2,\omega_2)=
W_1^{(e)}(t_1,\omega_1)\,W^{(e)}_2(t_2,\omega_2)$ as expected for
uncorrelated classical objects\footnote{Here the channel index breaks the
indiscernability between electrons within the two different channels.}. 
Consequently, if
the two-electronic Wigner distribution in channels $1$ and $2$ are
classical, the inter-channel two-electron Wigner distribution is also
classical.
But as we shall see now, because of its antisymmetry properties,
the two-electron Wigner distribution in a single channel exhibits non classical
features. 

To illustrate this point, let us consider mutually orthogonal
time-shifted wave-packets: $\varphi_1(t)=\varphi_e(t-\tau/2)$ and
$\varphi_2(t)=\varphi_e(t+\tau/2)$. 
The intrinsic two-electron
Wigner distribution function associated with the state $|\Psi_2\rangle=
\psi^\dagger[\varphi_1]\psi^\dagger[\varphi_2]|F\rangle$ 
is then
\begin{align}
\label{eq:G2:Wigner:2e-example}
\Delta & W^{(2e)}_{|\Psi_2\rangle} (t_1,\omega_1;t_2,\omega_2) = \notag \\
&W_{\varphi_1}\left(t_1,\omega_1\right)
W_{\varphi_2}\left(t_2,\omega_2\right)\notag \\
+ & W_{\varphi_2}\left(t_1,\omega_1\right)
W_{\varphi_1}\left(t_2,\omega_2\right)\notag \\
- & 2\cos{((\omega_1-\omega_2)\tau)}\,W_{\varphi_e}(t_1,\omega_1)W_{\varphi_e}(t_2,\omega_2)\,.
\end{align}
When considering a quasi-classical electronic wavepacket, such that
$W_{\varphi_e}(t,\omega)$ is almost everywhere positive, we see that the
last term contains interference fringes due to the
$\cos{((\omega_1-\omega_2)\tau)}$ factor. When 
$\varphi_1$ and $\varphi_2$ are well
separated, negativities appear which reflect the
non-classical nature of two-electron wavefunctions within a single edge
channel. Note that the dependence in $\omega_1-\omega_2$ comes from
the fact that, in the above example, 
$\varphi_{1}$ and $\varphi_2$ are time-shifted wavepackets.
Energy-shifted wavepackets would lead to
oscillations in $t_1-t_2$ analogous to Friedel oscillations in 
solid-state physics. In full generality, the quantum exchange interference
terms present both a
time and an energy dependence and this prevents $W^{(2e)}$ to be
interpreted as a time-dependent two-electron distribution function.

Similarly, the two-electron Wigner distribution function of
the equilibrium state at electronic temperature $T_{\text{e}}$ and 
vanishing chemical potential is given by
\begin{subequations}
\label{eq:W2:Fermi}
\begin{align}
\label{eq:W2:Fermi:1}
W^{(2e)}_{\mu=0,T_{\text{e}}}&(t_1,\omega_1;t_2,\omega_2)=
f_{T_{\text{e}}}(\omega_1)\,f_{T_{\text{e}}}(\omega_2) \\
\label{eq:W2:Fermi:2}
-4\pi k_B&T_{\text{e}}\delta(\omega_1-\omega_2)
f_{B,T_{\text{e}}}(\omega_\text{tot})\,
\frac{\sin{\left(\omega_\text{tot}t_{12}\right)}}{\sinh{\left(t_{12}/\tau(T_{\text{e}})\right)}}\,
\end{align}
\end{subequations}
where 
$\omega_\text{tot}=\omega_1+\omega_2$, $t_{12}=t_1-t_2$ and
$\tau(T)=\hbar/k_BT$ denotes the thermal coherence time.
Here
$f_{T_{\text{e}}}$ is the Fermi-Dirac distribution at temperature
$T_{\text{e}}$ and $\mu=0$ whereas
$f_{B,T}(\omega)=1/(\me^{\hbar\omega/k_BT}-1)$ denotes the Bose-Einstein
distribution at temperature $T$. The singular second term \eqref{eq:W2:Fermi:2}
expresses the Pauli exclusion principle and presents strong oscillations in
$t_{12}$. Therefore $W^{(2e)}_{\mu=0,T_{\text{e}}}$ cannot be interpreted as a 
time-dependent
electronic distribution.

\subsection{Relation to current noise} 
\label{sec:G2:noise}

Let us now describe the precise relation between two-electron
coherence and the excess current noise $\Delta S_i(t,t')$
defined as the excess of
\begin{equation}
\label{eq:G2:noise:definition}
S_i(t,t')=\langle i(t')\,i(t)\rangle - \langle i(t)\rangle\langle
i(t')\rangle\,.
\end{equation}
Since sub-nanosecond time-resolved measurements are not
available in electronics, $S_i(t,t')$ is not directly
accessible. However, 
finite-frequency current noise measurements~\cite{Parmentier:2010-1} 
give access to the noise
spectrum which is a time-averaged quantity. More recently, 
partial measurements of the time-dependent current noise
power spectrum have been performed. This quantity is  defined as the
Wigner function associated with excess current correlations:
\begin{equation}
\label{eq:G2:noise:Wigner-Ville}
W_{\Delta S_i}(t,\omega)=\int_{\mathbb{R}} \Delta S_i\left(t-\frac{\tau}{2}\Big\vert
t+\frac{\tau}{2}\right)\,\me^{\mi\omega\tau}\,\md\tau\,.
\end{equation}
This is achieved through a recently developped homodyne measurement
technique~\cite{Gasse:2013-1}
which has been used to probe the squeezing of the
radiation emitted by a tunnel junction.
The canonical anticommutation relations and definition
\eqref{eq:G2:decomposition} imply that the 
quantity defined by Eq.~\eqref{eq:G2:noise:Wigner-Ville}
is directly related to the intrinsic two-electron coherence by
\begin{subequations}
\label{eq:G2:noise:noise-to-G2}
\begin{align}
& W_{\Delta S_i}(t,\omega)+W_{\langle i\rangle}(t,\omega)= \notag \\
&-e\langle i(t)\rangle \label{eq:G2:noise:noise-to-G2:Poisson}\\
\label{eq:G2:noise:noise-to-G2:HBT}
&- e^2\int_{\mathbb{R}} h_\mu(\omega,\omega')\,\Delta
W^{(e)}_{\rho}(t,\omega')\,\frac{\md\omega'}{2\pi}\\
\label{eq:G2:noise:noise-to-G2:G2}
&+e^2\int_{\mathbb{R}}
v_F^2\Delta\mathcal{G}^{(2e)}_\rho\left(t+\frac{\tau}{2},t-\frac{\tau}{2}\Big\vert
t+\frac{\tau}{2},t-\frac{\tau}{2}\right)\,\me^{\mi\omega\tau}\md\tau\,
\end{align}
\end{subequations}
where
$h_{\mu}(\omega,\omega')=f_{\mu}(\omega-\omega')+f_{\mu}(\omega+\omega')$
and $W_{\langle i\rangle}(t,\omega)$ denotes the Wigner-Ville
transform~\cite{Ville:1948-1}
of the average time-dependent current.
The first term~\eqref{eq:G2:noise:noise-to-G2:Poisson} is a
Poissonian contribution associated with the granular nature of charge
carriers. 
The second term \eqref{eq:G2:noise:noise-to-G2:HBT} arises from 
two-particle interferences between excitations generated by the source and electrons
within the Fermi sea. These contributions are called the HBT
contributions since these two-particle interferences are precisely what
is measured in an HBT experiment.
Finally, the last term
\eqref{eq:G2:noise:noise-to-G2:G2} corresponds to the intrinsic two-electron
coherence contribution to the current noise. This equation is indeed the
electron quantum optics version of the famous relation on fluctuations
of particle number in an ideal Bose gas~\cite{Einstein:1925-1} where the
term \eqref{eq:G2:noise:noise-to-G2:G2} 
minus $W_{\langle i\rangle}(t,\omega)$ corresponds to
Einstein's quadratic term.
Finally, the non triviality of the Fermi sea vacuum is responsible for the term
\eqref{eq:G2:noise:noise-to-G2:HBT}.

This equation also relates electronic coherences to the quantum optical
properties of the edge magnetoplasmons within the edge channel and
therefore of the photons radiated into a transmission line capacitively
coupled to the edge channel as 
in \cite{Degio:2009-1}. It therefore establishes a bridge between
electron quantum optics and the recently studied quantum optics of
noise~\cite{Grimsmo:2016-1}.

Finally, let us stress that Eq.~\eqref{eq:G2:noise:noise-to-G2}, which is also
valid in the presence of
interactions, shows
that accessing single-electron coherence as well as the
current noise gives access to the diagonal part of two-electron
coherence, as expected since the latter contains all the
information on time resolved two-electron detection.

Directly accessing the intrinsic two-electron coherence without any
HBT contribution can be achieved by
partitioning the electronic beam onto a beam splitter in an HBT setup
(see Fig.~\ref{fig:HBT-HOM:principle}). 
These current correlations are directly related to the 
inter-channel two-electron coherence right after the beam splitter 
since fermionic fields within different channels anticommute:
\begin{align}
\label{eq:G2:HBT:current-correlations}
\langle i_{1\mathrm{out}}(t_1)&i_{2\mathrm{out}}(t_2)\rangle = \notag \\
& e^2v_F^2 \Delta\mathcal{G}^{(2e)}_{\mathrm{out}}(1,t_1;2,t_2|1,t'_1;2,t'_2)
\end{align}
Remarkably, this outgoing two-electron coherence is proportional to
the incoming two-electron coherence~\cite{Thibierge:2016-1}:
\begin{align}
\label{eq:G2:HBT:1source}
\Delta\mathcal{G}_{\mathrm{out}}^{(2e)}(1,t_1;2,t_2&|1,t'_1;2,t'_2)= \notag \\
&RT\,\Delta\mathcal{G}^{(2e)}_S(t_1,t_2|t'_1,t'_2)\,.
\end{align}
Measuring outgoing inter-channel current correlations in the HBT setup
thus directly probes the intrinsic excess coherence of the source, as
in photon quantum optics.

\subsection{Two-particle interferometry as linear filtering} 
\label{sec:G2:Franson}

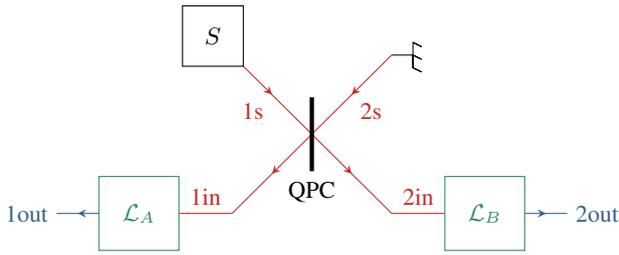
\begin{figure}
\begin{tikzpicture}[scale=0.7,
  decoration={
    markings,% switch on markings
    mark=at position .5 with {\arrow{stealth}}}
  ]
% Nommage de points au centre des beam splitters
\coordinate (O) at (0,0);
\coordinate (entree_MZI1) at (-2.5,-1.5);
\coordinate (sortie_MZI1) at (-4,-1.5);
\coordinate (entree_MZI2) at (2.5,-1.5);
\coordinate (sortie_MZI2) at (4,-1.5);

\coordinate (source1) at (-1.5,1.5);
\coordinate (source2) at (1.5,1.5);

% Canaux du secteur SOURCE
\draw [myred, postaction={decorate}] (source2) -- (O) 
  node [midway, below right] {2s};
\draw [myred, postaction={decorate}] (source1) -- (O)
  node [midway, below left] {1s};

\begin{scope}[xshift = 1.5cm, yshift=1.5cm, rotate=90, scale=0.4]
\draw (0,0) -- ++(0,-1)
	-- ++(-.6,0) -- ++(-.2,-.4) -- ++(.2,.4)
	-- ++(.6,0) -- ++(-.2,-.4) -- ++(.2,.4)
	-- ++(.6,0) -- ++(-.2,-.4) -- ++(.2,.4);
\end{scope}

\node[at={($(source1)+(-.36cm,.36cm)$)}] (source){
  \begin{tikzpicture}
    \begin{scope}[color= black]
      \filldraw [fill = white] (0,0) node {$S$}  (-0.4,-0.4) rectangle (0.4,0.4);
    \end{scope}
  \end{tikzpicture}
};

% Canaux du secteur IN
\draw [myred,postaction={decorate}] (O) -- ++(-1.5,-1.5);
\draw [myred] (-1.5,-1.5) -- (entree_MZI1)
  node [midway, above] {1in};
\draw [myred, postaction={decorate}] (O) -- ++(1.5,-1.5);
\draw [myred] (1.5,-1.5) -- (entree_MZI2)
  node [midway, above] {2in};

% Blocs linéaires
\draw [mygreen] (entree_MZI1)-- ++ (0,-0.7) -- ++(-1.5,0) -- ++(0,1.4) -- ++ (1.5,0) --cycle;
\draw [mygreen] (entree_MZI2)-- ++ (0,-0.7) -- ++(1.5,0) -- ++(0,1.4) -- ++ (-1.5,0) --cycle;

\draw [mygreen] (entree_MZI1) ++(-0.75,0) node {$\mathcal{L}_{A}$};
\draw [mygreen] (entree_MZI2) ++(0.75,0) node {$\mathcal{L}_{B}$};

% Canaux du secteur OUT
\draw [myblue, postaction={decorate}] (sortie_MZI1) -- ++(-0.8,0)
  node [left] {1out};
\draw [myblue, postaction={decorate}] (sortie_MZI2) -- ++(0.8,0)
  node [right] {2out};

% Beam splitter
\draw [ultra thick] (O) ++(0,.7) -- ++(0,-1.4) node [below] {QPC};
\end{tikzpicture}

\caption{\label{fig:Franson-generalized} A generalized Franson
interferometry experiment: the electron flow emitted by the source $S$ is
partitionned at the QPC and sent into two linear filtering
components $A$ and $B$. Current correlations between outgoing currents
give access to second-order electronic coherence. Franson
interferometry corresponds to using two
Mach-Zehnder interferometers for $A$ and $B$.}
\end{figure}

Although current correlations in the HBT geometry only give access
to the diagonal part of the intrinsic two-electron coherence in the 
time domain, Eq.~\eqref{eq:G2:HBT:current-correlations} naturally leads to
a general idea for accessing the off-diagonal part
$\Delta\mathcal{G}_S^{(2e)}(t_1,t_2|t'_1,t'_2)$ for $(t_1,t_2)\neq
(t'_1,t'_2)$. The idea is to use linear filters 
of single-electron coherence as depicted on
Fig.~\ref{fig:Franson-generalized}. Let us assume that the outgoing current
is obtained from a linear filtering of the incoming single-electron
coherence $\langle
i_A\rangle=\mathcal{L}_A\left[\Delta\mathcal{G}_{1\mathrm{in}}^{(e)}\right]
$ with a similar relation for detector $B$. Then, 
the outgoing current correlations
$\langle i_A\,i_B\rangle$ are obtained by applying a linear filter to the
incoming two-electron coherence:
\begin{equation}
\label{eq:G2:Franson:main}
\langle i_A\, i_B\rangle 
= RT\left(\mathcal{L}_A^{(1)}\otimes
\mathcal{L}_B^{(2)}\right)\left[\Delta\mathcal{G}^{(2e)}_S\right]
\end{equation}
in which \eqref{eq:G2:HBT:1source} has been used to obtain
\eqref{eq:G2:Franson:main}.

Despite its compacity, Eq.~\eqref{eq:G2:Franson:main} unifies many
different experiments
under a simple physical interpretation: the intrinsic two-electron
coherence $\Delta\mathcal{G}_S^{(2e)}$ describing two-particle excitations
emitted by the source, is encoded into current correlations $\langle
i_A\,i_B\rangle$ via an HBT interferometer 
and two linear filters for single-electron coherence. 

In the absence of 
these filters, the measurement of current correlation
gives information on the diagonal part of $\Delta\mathcal{G}_S^{(2e)}$ as
seen in section \ref{sec:G2:noise}. When $A$ and $B$ are electronic energy filters, and
assuming that no electronic relaxation process takes place between the
QPC and the filters (see the discussion in section \ref{sec:G1:MZI}), we access the diagonal
part of $\Delta\mathcal{G}^{(2e)}_S$ in the frequency domain. 

It also naturally leads to the idea of 
the Franson interferometer~\cite{Franson:1989-1} originally invented to test
photon
entanglement~\cite{Brendel:1999-1,Marcikic:2002-1} and later considered
for
testing two-particle Aharonov-Bohm effect and
electronic entanglement generation~\cite{Splettstoesser:2009-1}. It is a natural way to
probe the off-diagonal part of $\Delta\mathcal{G}_S^{(2e)}$ in the time
domain since, as explained in section \ref{sec:G1:MZI}, a MZI converts
single-electron coherences in the time domain into electrical
currents. 

\section{From electron quantum optics to quantum information}
\label{sec:QI}

In this last section, we explain how to relate electron quantum optics
quantities to quantum information ones using, once again, signal
processing ideas. This enables discussing quantum information
quantities in electron
quantum optics systems. We illustrate these ideas and procedures by 
sketching how to get an insight on the many-body state generated by the
mesoscopic capacitor.

\subsection{Density matrices from electronic coherences}
\label{sec:QI:density-matrices}

Because electrons are fermions, each single-particle state can be viewed
as a two-level system, thus raising the question of their use as effective qubits
to encode quantum information. Given a normalized 
wavefunction $\varphi_a$, its average occupation number is obtained
from single-electron coherence by
\begin{equation}
\label{eq:QI:population}
\bar{n}[\varphi_a]=
v_F^2\int_{\mathbb{R}^2}
\varphi_a(t_+)^*\varphi_a(t_-)\,\mathcal{G}^{(e)}_{\rho,x}(t_+|t_-)\,
\md t_+\,\md t_-\,
\end{equation}
where $\varphi_a(t)$ denotes the electronic wavefunction for $x=-v_Ft$. 
The quantity $\bar{n}[\varphi_a]$ being the average value of
the number operator $N[\varphi_a]=\psi^\dagger[\varphi_a]\psi[\varphi_a]$,
it is between zero and unity. 
Denoting by $|0_a\rangle$ and $|1_a\rangle$ the states respectively
corresponding to the absence or the presence of an electron in the single
particle state $\varphi_a$, $\bar{n}_a$ and $1-\bar{n}_a$ 
can be used as diagonal elements of a $2\times 2$ matrix.
However, in the absence of
superconductors, charge conservation leads to a superselection rule 
forbidding quantum superpositions of states of different charges:
its off-diagonal elements, which are equal to $\langle
\psi^\dagger[\varphi_a]\rangle_\rho$ and its complex conjugate,
vanish. Nevertheless, this idea becomes more interesting when considering
more than one single-particle state because the various superselection
sectors are no longer one dimensional. 

A first possibility to obtain an effective qubit is to consider
two orthogonal single-particle states $\varphi_a$ and $\varphi_b$ and the charge one sector. 
The basis vector $|0\rangle$ (resp. $|1\rangle$) then corresponds to the state where the
electron is totally localized in the electronic state $\varphi_a$ (resp.
$\varphi_b$).
The reduced
density matrix for this railroad electronic
qubit~\cite{Ionicioiu:2001-1} is then defined from single and two
electron coherences by:
\begin{subequations}
\begin{align}
\langle 0|\rho_{\text{qb}}&|0\rangle= 
\langle
\psi^\dagger[\varphi_a]\psi[\varphi_a]\,\psi[\varphi_b]\psi^\dagger[\varphi_b]\rangle_\rho\\
\langle 1|\rho_{\text{qb}}&|1\rangle
=\langle
\psi^\dagger[\varphi_b]\psi[\varphi_b]\,\psi[\varphi_a]\psi^\dagger[\varphi_a]\rangle_\rho
\\
\langle 0|\rho_{\text{qb}}&|1\rangle=
\langle \psi^\dagger[\varphi_a]\psi[\varphi_b]\rangle_\rho
\end{align}
\end{subequations}
This framework can be extended to situations involving
more single-particle states. For example, one could consider two pairs
of single-electron states $\varphi_{a1}$, $\varphi_{a2}$ and
$\varphi_{b1}$ $\varphi_{b2}$ as in the four leads device
considered by Samuelsson and B\"{u}ttiker \cite{Samuelsson:2006-1}. 
The two-particle sector would then contain an
effective 2-qubit reduced density matrix whose matrix elements can be
expressed in terms of electronic coherences.

Deriving an effective two-qubit reduced density operator
from electron coherences
directly enables us to use the results from quantum information on the
characterization of entanglement in a bipartite system. 

Entanglement is well defined and easily characterized for bipartite systems in a pure
state using for example the von
Neumann entropy of
the reduced density matrix of one of the two
subsystems~\cite{Book:Nielsen-Chuang}. In cases where the
whole system is not in a pure state, its total density matrix is said to
represent an entangled state if and only if it is not factorized, that
is, if and only if it cannot be written as a
statistical mixture of tensor products of density operators relative to
each subsystem. In the case of a two qubit system, a single quantity
called the concurrence can be used to characterize entanglement.
Denoting by $\rho_{\text{2qb}}$ the total density matrix for the two
qubits, the concurrence is defined by~\cite{Hill:1997-1}:
\begin{equation}
C[\rho_{\text{2qb}}]=\max{(0,\lambda_1-\lambda_2-\lambda_3-\lambda_4)}
\end{equation}
where $\lambda_1\geq \lambda_2\geq \lambda_3\geq \lambda_4$ are the
eigenvalues of the hermitian matrix
$R=\sqrt{\sqrt{\rho_{\text{2qb}}}\cdot
\tilde{\rho}_{\text{2qb}}\cdot\sqrt{\rho_{\text{2qb}}}}$, and 
\begin{equation}
\tilde{\rho}_{\text{2qb}}=(\sigma^y\otimes \sigma^y)\,\rho_{\text{2qb}}^*\,
(\sigma^y\otimes \sigma^y)\,.
\end{equation}
The concurrence vanishes if and only if the 2-qubit state is factorized.

The entanglement of
formation represents the minimum of the average entanglement of an
ensemble of pure states representing
$\rho_{\text{2qb}}$~\cite{Wootters:1998-1}. It
corresponds to the minimal number of maximally entangled qubits per
realization\footnote{As often in quantum information, 
we are looking for the number of maximally entangled pairs needed to
generate $n$ copies of the states contained in $\rho$. For large $n$,
this scales as $n$ and we divide by
$n$ to obtain the entanglement of formation.} 
needed to generate the entangled states described by
$\rho_{\text{2qb}}$~\cite{Bennett:1996-1,Wootters:2001-1}. Remarkably,
for a 2-qubit, it has a closed expression in terms of 
the concurrence~\cite{Wootters:1998-1}:
\begin{subequations}
\begin{align}
\mathcal{E}[\rho_{\text{2qb}}]&=H_2\left(\frac{1+\sqrt{1-\mathcal{C}[\rho_{\text{2qb}}]^2}}{2}\right)\\
H_2(x)& =-x\log_2{(x)}-(1-x)\log_2{(1-x)}
\end{align}
\end{subequations}
These considerations describe how to obtain quantum information
quantities from electron quantum optics concepts. 
Let us now illustrate these ideas on the study of electron/hole entanglement
generated by the mesoscopic capacitor.

\subsection{Electron/hole entanglement from the mesoscopic capacitor}
\label{sec:LPA-source}

\begin{figure}
\begin{center}
 \begin{tikzpicture}

%Zone du 2DEG
\draw[fill=green!10!white] (1,0.7) -- (6.5,0.7) -- (6.5,2) -- (1,2) -- cycle;
\draw[fill=green!10!white,rounded corners = .2cm] 
(4.2,2) -- (3.9,2.4) -- %
(4,2.8) arc [start angle=-60,   end angle=240, x radius=1cm, y radius=1cm] 
-- (3.1,2.4) -- (2.8,2);

%QPC
\draw[fill=cyan!20!white] (2.2,2.1) -- (2.9,2.3) --(2.9,2.5) -- (2.2,2.7) -- cycle;
\draw[fill=cyan!20!white] (4.8,2.1) -- (4.1,2.3) --(4.1,2.5) -- (4.8,2.7) -- cycle;

\draw[thick, densely dotted, rounded corners = .15cm] (2.9,1.7) -- (3.2,2.4) -- (3,3.1); 
\draw[thick, densely dotted, rounded corners = .15cm] (4.1,1.7) -- (3.8,2.4) -- (4,3.1); 
\draw[<->] (3.35,2.1) -- (3.35,2.7) node[midway,right]{$D$};

%Canaux de bords
\begin{scope}[thick, decoration={
    markings,
    mark=at position 0.2 with {\arrow{stealth}},
    mark=at position 0.75 with {\arrow{stealth}}}
    ] 
\draw[color = red, postaction={decorate}] (1,1.3) -- (6.5,1.3);
\end{scope}

\begin{scope}[thick, decoration={
    markings,
    mark=at position 0.5 with {\arrow{stealth}}}
    ] 
\draw[color = blue, postaction={decorate}] (1,1.7) -- (2.9,1.7);
\draw[color = blue, postaction={decorate}] (4.1,1.7) -- (6.5,1.7);
\end{scope}

\draw[thick, densely dotted] (2.9,1.7) -- (4.1,1.7);
\draw[->] (3.3,1.85) -- (3.7,1.85);

\begin{scope}[thick, decoration={
    markings,
    mark=at position 0.3 with {\arrow{stealth reversed}},
    mark=at position 0.7 with {\arrow{stealth reversed}}}
    ] 
  \draw[color = blue, postaction={decorate}] (3.5,3.7) circle (0.75);
\draw[color = red, postaction={decorate}] (3.5,3.7) circle (0.5);  
\end{scope}

%Top Gate
\draw[fill=pink,opacity=0.4] (4.5,2.8) -- (4.7,4.7) -- (2.3,4.7) -- (2.5,2.8) -- cycle;

\draw (3.5,4.7) -- (3.5,5) -- (6,5) -- (6,4.5);
\draw (5.7,4.5) -- (6.3,4.5);
\draw (5.7,4.35) -- (5.8,4.5);
\draw (5.85,4.35) -- (5.95,4.5);
\draw(6,4.35) -- (6.1,4.5);
\draw (6.15,4.35) -- (6.25,4.5);

\draw[fill=white] (5.2,5) circle (0.45);
\node at (5.2,5) {$V_d(t)$};

\node at (3.5,1.5) {$1-D$};

 \end{tikzpicture}
\end{center}
\caption{\label{fig:source-LPA} The mesoscopic capacitor consists of a
quantum dot with level spacing $\Delta$ connected by a QPC (in light blue) to an edge channel (here the 
outer one in a $\nu=2$ system). $D$ denotes the
transmission probability of the dot so that $D=1$ corresponds to the
case where the dot is totally open and $D=0$ to the case where the dot
is disconnected from the edge channel.
Electrons within the dot experience the time-dependent drive potential 
$V_d(t)$ imposed by a top gate (in light pink).
For a square ac voltage at frequency $f$ 
with zero average and amplitude $\Delta/e$, there exist 
an optimal value $D_{\mathrm{opt}}$ function of $\Delta/hf$ for single-electron and 
hole emission.
}
\end{figure}
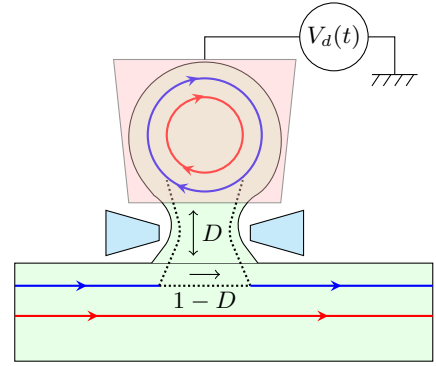

The mesoscopic capacitor depicted on Fig.~\ref{fig:source-LPA} is a
quantum RC circuit~\cite{Gabelli:2006-1} which can be operated in the 
non-linear regime with a periodic square drive at frequency $f=1/T$
in order to emit a single-electron excitation
during the half period $[0,T/2]$ and a single hole excitation during the
second half period $[T/2,T]$~\cite{Feve:2007-1}. Reaching this optimal single-electron
source regime requires tuning the QPC transparency $D$ (see
Fig.~\ref{fig:source-LPA}) so that the electron and hole excitations
have the time to form and escape during their respective half periods.

The source is modeled as a time-dependent single-electron scatterer
which amounts to neglecting the effect of Coulomb interactions within the
dot. Under this hypothesis, the single-electron coherence can be
computed using Floquet's theory~\cite{Moskalets:book,Moskalets:2002-1}.
Note that in this case, single-electron
coherence determines
all the electronic coherences emitted by the source since Wick's theorem
is valid. Here we
use a Floquet description of \cite{Degio:2010-4} to
obtain the single-electron coherence generated by the mesoscopic
capacitor. A program written in C computes the single-electron coherence
in the frequency domain. An Haskell code then computes 
2-qubit matrices and quantum information theoretical quantities from
these data.

We have considered the mesoscopic capacitor with the following
experimentally relevant parameters: $\Delta/hf =20$.
The drive voltage $V_d(t)$ is a square voltage.
The optimal operation point with
these parameters is found to be at $D_{\text{opt}}\simeq
0.35$. 

\begin{figure*}
%\missingfigure{Plots de Wigner}
\begin{center}
\includegraphics{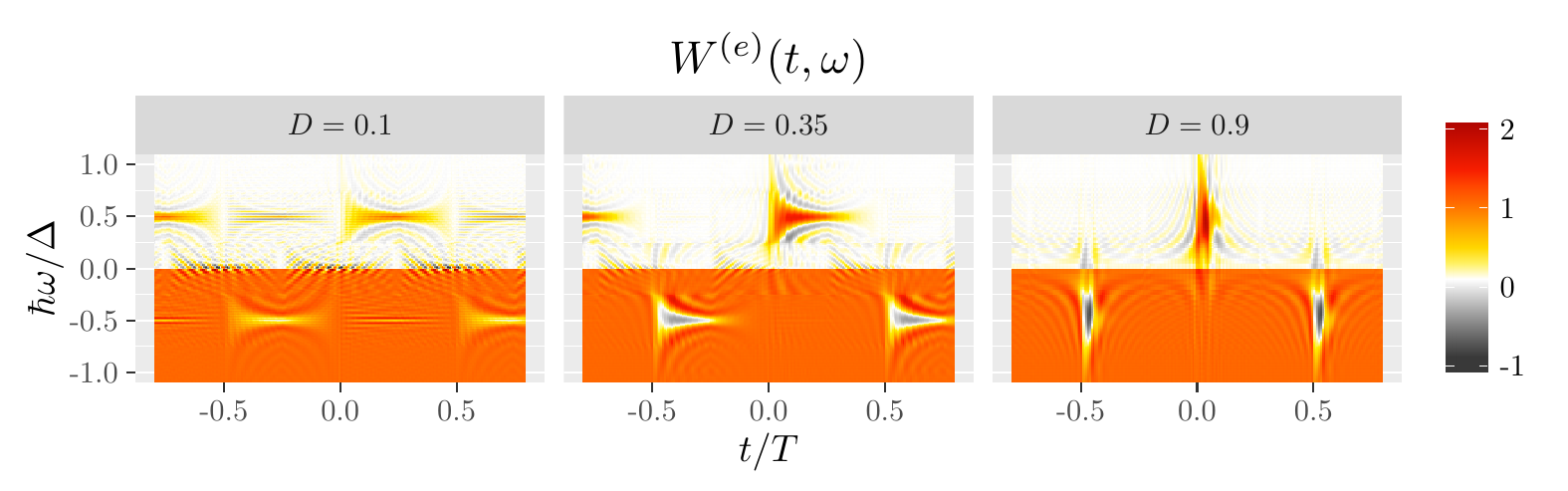}
\end{center}
\caption{\label{fig:source-LPA:We} The electronic Wigner function
$W^{(e)}(t,\omega)$ for the mesoscopic capacitor operated at the
electron/hole symmetric point and driven by a square potential of
amplitude $\Delta/e$ at frequency $f$ such that $\Delta/hf=20$. Plots
are presented for
$D=0.1$, $D=D_{\text{opt}}\simeq 0.35$ and $D=0.9$.
}
\end{figure*}

Figure \ref{fig:source-LPA:We} presents the electronic
Wigner distribution defined by Eq. \eqref{eq:G1:Wigner} for $D=0.9$,
$D\simeq D_{\mathrm{opt}}$ and $D=0.1$. These plots clearly show that when the
dot is fully open, energy resolution is lost, whereas at
$D\simeq D_\text{opt}$, we clearly see the single-electron and single-hole
excitations emitted during each half period. Decreasing $D$ leads to two
phenomena: first of all, the length of the electronic wavepacket
increases and we see horizontal interference fringes 
between two-electronic emissions which suggests delocalization of the
electronic excitation beyond the time interval $[0,T]$
(same for the hole). We also see interference fringes between the
electronic and hole contributions which correspond to an increase of the
weight of $\Delta\mathcal{G}^{(e)}_S$ in the electron/hole coherence
quadrant in the frequency domain~\cite{Degio:2010-1}.
It was suggested in \cite{Degio:2010-4} that, in this regime, during each
period, the source emits a state of the form
$|\Psi_\text{e/h}(u,v)\rangle=(u+v\,\psi[\varphi_h]\psi^\dagger[\varphi_e])|F\rangle$ where
$\varphi_e$ and $\varphi_h$ respectively denote the electronic and hole
excitations and $(u,v)$ complex amplitudes such that $|u|^2+|v|^2=1$. At the optimal point
$(u,v)\simeq (0,1)$ whereas $(u,v)\simeq (1,0)$ when $D\to 0^+$ since
the source is shut down in this limit. Consequently, we have
$|u|^2\simeq |v|^2\simeq 1/2$ for some intermediate value.
In the latter case, this state contains a quantum superposition between the presence
and the absence of an elementary electron/hole pair excitation in the edge channel. 

Elaborating on section \ref{sec:QI:density-matrices}, we will now 
test this idea by quantifying the amount of
entanglement for an effective 2-qubit built from an electronic and a
hole excitation. 
For this purpose, we define a $4\times 4$ matrix $\rho_{\text{e/h}}$ built
from states $|x_h\,x_e\rangle$ associated with the occupation
number $x_h\in \{0,1\}$ for a hole excitation based on the single-particle
wavefunction $\varphi_h$ and the electronic occupation number
$x_e\in \{0,1\}$ 
associated with the single-particle state $\varphi_e$. All
matrix elements that couple different charge sectors vanish. The
remaining matrix elements are the diagonal ones:
\begin{subequations}
\label{eq:ehqb:diagonal}
\begin{align}
\langle 00|\rho_{\text{e/h}}&|00\rangle= 
\langle
\psi^\dagger[\varphi_h]\psi[\varphi_h]\,
\psi[\varphi_e]\psi^\dagger[\varphi_e]
\rangle_\rho\\
\langle 01|\rho_{\text{e/h}}&|01\rangle=
\langle
\psi^\dagger[\varphi_h]\psi[\varphi_h]\,
\psi^\dagger[\varphi_e]\psi[\varphi_e]
\rangle_\rho
\\
\langle 10|\rho_{\text{e/h}}&|10\rangle=
\langle 
\psi[\varphi_h]\psi^\dagger[\varphi_h]\,
\psi[\varphi_e]\psi^\dagger[\varphi_e]
\rangle_\rho
\\
\langle 11|\rho_{\text{e/h}}&|11\rangle=
\langle
\psi[\varphi_h]\psi^\dagger[\varphi_h]\,
\psi^\dagger[\varphi_e]\psi[\varphi_e]
\rangle_\rho
\end{align}
\end{subequations}
and the off-diagonal elements coupling the state
$|00\rangle$ to the state
$|11\rangle$:
\begin{equation}
\label{ehqb:non-diagonal:1}
\langle 11|\rho_{\text{e/h}}|00\rangle=
\langle \psi^\dagger[\varphi_h]\psi[\varphi_e]\rangle_S\,.
\end{equation}
The diagonal matrix elements \eqref{eq:ehqb:diagonal} are all related to
the two-electron coherence
$\bar{n}[\varphi_e,\varphi_h]$
whereas
the off-diagonal ones are directly the single-electron coherence in the
electron/hole coherence quadrant
$\mathcal{G}_S^{(e)}[\varphi_e|\varphi_h]=
\mathrm{Tr}(\psi[\varphi_e]\,\rho\,\psi^\dagger[\varphi_h])$.

When Wick's theorem is valid, which is the case in our Floquet
modelization of the mesoscopic capacitor, we have:
\begin{subequations}
\label{eq:ehqb:diagonal:2}
\begin{align}
\langle 00|\rho_{\text{e/h}}|00\rangle &= 
(1-\bar{n}_e)\bar{n}_h 
+
|\xi|^2
\\
\langle 01|\rho_{\text{e/h}}|01\rangle &=
\bar{n}_e\bar{n}_h 
-
|\xi|^2
\\
\langle 10|\rho_{\text{e/h}}|10\rangle &=
(1-\bar{n}_e)(1-\bar{n}_h) 
-
|\xi|^2
\\
\langle 11|\rho_{\text{e/h}}|11\rangle &=
\bar{n}_e(1-\bar{n}_h) 
+
|\xi|^2
\end{align}
\end{subequations}
where $\xi=\left\langle 11 \middle| \rho_{\text{e/h}} 
\middle| 00\right\rangle=\mathcal{G}_S^{(e)}[\varphi_e|\varphi_h]$ is the
electron/hole coherence between the electronic and hole wavefunctions
and $\bar{n}_{e/h}=\bar{n}[\varphi_{e/h}]$ are their
occupation numbers. The Cauchy-Schwarz inequality tells us that
\begin{equation}
\label{eq:G1:CS}
\left| \mathcal{G}_S^{(e)}[\varphi_e|\varphi_h]\right|^2\leq
\bar{n}_e\,\bar{n}_h\,.
\end{equation}
When the bound is saturated, the matrix $\rho_\text{e/h}$ gets a zero on
the diagonal and, in this case, the concurrence can be evaluated
analytically:
\begin{equation}
\label{eq:QI:C-analytical}
\mathcal{C}[\rho_\text{e/h}]=2\sqrt{\bar{n}_e\,\bar{n}_h}\,.
\end{equation}
It is non zero when we are away from the single-electron source
regime ($\bar{n}_e=1$ and $\bar{n}_h=0$). In particular for
the state $|\Psi_\text{e/h}(u,v)\rangle$ which can be shown to 
satisfy Wick's theorem, considering 
$|u|^2=|v|^2=1/2$ gives
$\bar{n}_e=\bar{n}_h=1/2$ and thus $\mathcal{C}=1$. Then,
$\mathcal{E}[\rho_\text{e/h}]=1$, as expected
since, in this case, we are producing a pure state with electron/hole
entanglement. 

However, for arbitratry electronic and hole wavepackets $\varphi_{e/h}$,
the reduced density operator $\rho_\text{e/h}$ may not be so ideal. We
shall now discuss which wavepackets are the best candidates for obtaining
non-zero concurrence.

\subsection{Numerical results}
\label{sec:QI:numerics}

As we shall see now, the choice of $\varphi_{e/h}$ has
a strong influence on the result. At fixed $\bar{n}_e$ and
$\bar{n}_h$, the concurrence is maximal for $\xi$ saturating the
Cauchy-Schwarz bound \eqref{eq:G1:CS} but its value can be very small
when $\bar{n}_e\simeq 0$ and $\bar{n}_h\simeq 1$. We thus have
to find wavepackets which (i) bring $\bar{n}_e$ and
$\bar{n}_h$ as close as possible to $1/2$ and
(ii) maximize $\vert\mathcal{G}_S^{(e)}[\varphi_e|\varphi_h]\vert$. 

A first guess for $\varphi_e$ is a truncated Lorentzian in energy, centered
at $\hbar\omega_e=\Delta/2$ and whose width fits the exponential decay
of the average current. Note that its natural width $\gamma_e$ depends
on $D$.
This wavepacket is expected to lead to $\bar{n}_e\simeq 1$ when
$D=D_\text{opt}$. In this case, the natural width of the wavepacket
should be less than a half period.
However, when $D\rightarrow 0$, it may be relevant to
consider electronic wavepackets delocalized over more than one
period. Our ansatz is to consider the product of the truncated
Lorentzian in energy with a characteristic function in time that selects 
$n$ half-periods during which the electron is emitted ($t\in
[lT,(l+1/2)T]$ for $l=0,\ldots,n-1$). 
Truncation in the time domain implies that our ansatz are not anymore
a purely electronic state. In order to ensure this, we set to zero all values at
negative frequencies. This
implies that the wavepackets we consider are not strictly zero in time outside
of $[lT, (l+1/2)T]$. However, when $D \le D_{\text{opt}}$ the excitations
are emitted at well separated energies $\hbar\omega_{e/h}=\pm \Delta/2$ with respect
to their natural width $\hbar\gamma_e$, so contributions outside these half
periods are very small (and indeed smaller as $D$ decreases). When $D \ge
D_{\text{opt}}$, the natural width of the excitation decreases, ensuring
it is less than a half period.
After normalization, this defines wavefunctions $\varphi_{e_n}$ and in
the same way wavefunctions $\varphi_{h_n}$.

\begin{figure*}
\begin{center}
\includegraphics{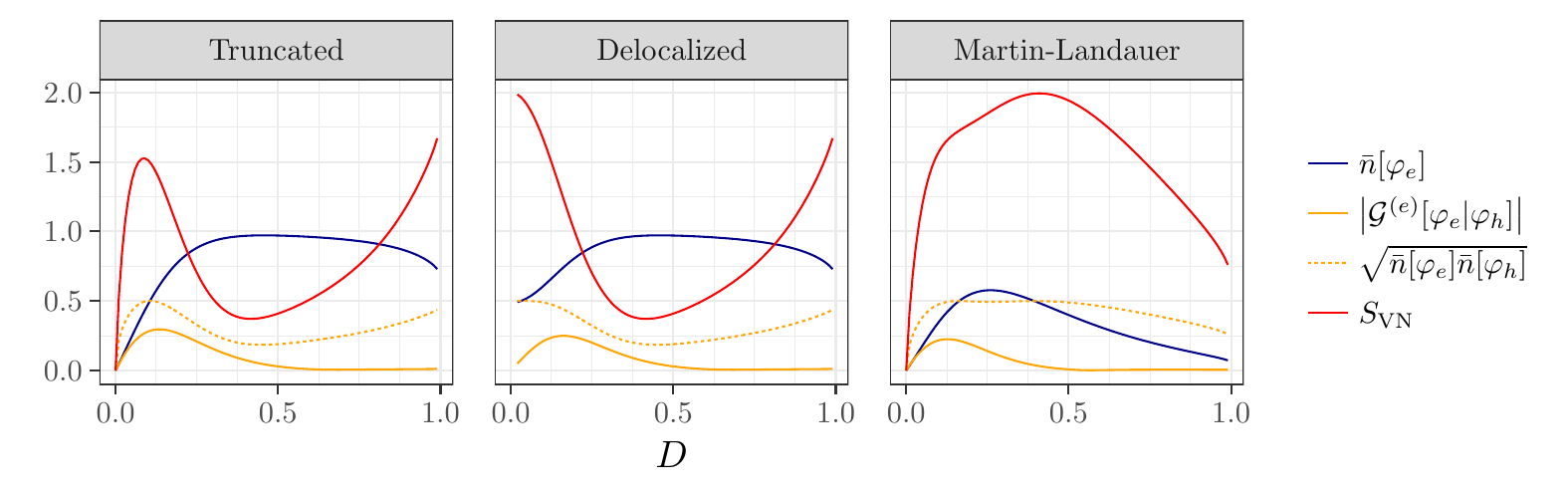}
\end{center}
\caption{\label{fig:source-LPA:ne-etc} 
The average occupation $\bar{n}_e$ (blue line), the modulus of the coherence
$\vert \mathcal{G}_S^{(e)}[\varphi_e|\varphi_h]\vert$ (orange line), the
Cauchy-Schwarz bound $\sqrt{\bar{n}_e\bar{n}_h}$ (dashed
orange line) and the von Neumann entropy $S_{\text{VN}}$ of $\rho_\text{e/h}$ (red
line) as functions of $D$ for the three families of wavepackets
``Truncated'', ``Delocalized'' and ``Martin Landauer''. The mesoscopic
capacitor is
operated with the parameters used for Fig.~\ref{fig:source-LPA:We}.
}
\end{figure*}

\begin{figure}
\begin{center}
\includegraphics{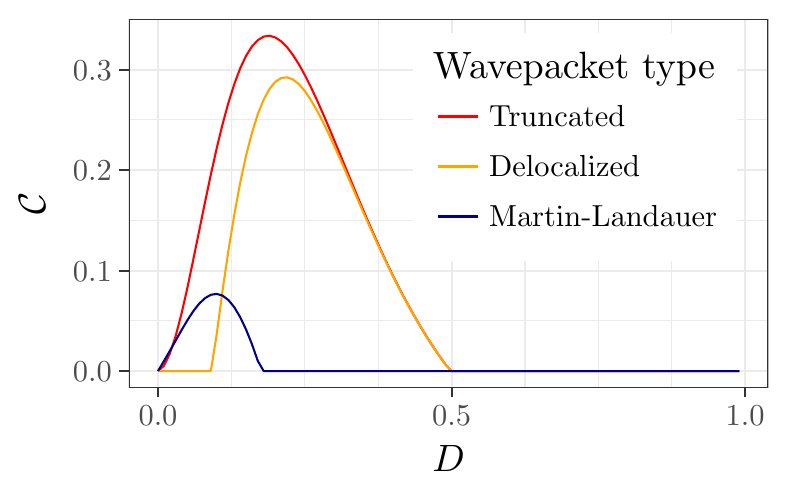}
\end{center}
\caption{\label{fig:source-LPA:concurrence} 
The concurrence $\mathcal{C}[\rho_{\text{e/h}}]$
as function of $D$ for the three families of wavepackets
``Truncated'', ``Delocalized'' and ``Martin Landauer''.
The mesoscopic capacitor is 
operated with the parameters used for Fig.~\ref{fig:source-LPA:We}.
}
\end{figure}

We present on Fig.~\ref{fig:source-LPA:ne-etc} the quantities
$\bar{n}_{e/h}$, the modulus of the electron/hole coherence $|\xi|$,
the Cauchy-Schwarz bound
and the von Neumann entropy of the 2-qubit matrix 
$\rho_{\text{e/h}}$ using specific couples of electronic and hole
wavefunctions. We consider
$\varphi_{e_n}$ and $\varphi_{h_n}$ for $n=1$, {\it id est} truncated to one half period
and thus denoted by ``Truncated'' on
Figs.~\ref{fig:source-LPA:ne-etc} and \ref{fig:source-LPA:concurrence}
and also $n=\infty$ which corresponds to the ``Delocalized''
wavefunctions over many half periods.

The von Neumann entropy of the 2-qubit density matrices obtained from
these two wavepacket families shows roughly the same behavior for
$D\gtrsim 0.1$.
First of all, it never vanishes for $D\neq 0$ thus showing that $\rho_\text{e/h}$ 
does not represent a rank one projector.
The von Neumann entropy goes through a minimum close to $D_\text{opt}$ 
and then re-increases. 
This comes from
the fact that these wavepackets are indeed well suited to describe the
excitations emitted by the mesoscopic capacitor around the optimal regime as
attested by the fact that $\bar{n}_e\simeq 1$ for this regime.

In the vanishing $D$ limit, $\bar{n}_e \approx 0.50$ for the delocalized
wavepacket whereas $\bar{n}_e\to 0$ for the
truncated ones. 
This reflects the delocalization over more than one half period of the
electronic excitations when $D\rightarrow 0$.
For the delocalized wavepackets, the Cauchy-Schwarz bound is
not saturated when $D\rightarrow 0$. The effective 2-qubit density
matrix $\rho_\text{e/h}$ thus becomes diagonal. Due to 
the non-vanishing limit of $\bar{n}_e$ when $D\rightarrow 0$,
the von Neumann entropy remains close to $2$ (which would
be obtained for $\xi=0$ and $\bar{n}_e=\bar{n}_h=1/2$). By
contrast, for the truncated wavepackets, $\bar{n}_e$ goes to $0$
and $\bar{n}_h\to 1$ when $D\rightarrow 0$. The Cauchy-Schwarz
bound is also not saturated. In this case, $\rho_\text{e/h}$ collapses onto a
matrix with only one non-zero element on its diagonal: the von Neumann
entropy decreases when $D\rightarrow 0$ for truncated wavepackets.

The concurrence as a function of $D$ is depicted on
Fig.~\ref{fig:source-LPA:concurrence}. We see that for delocalized
wavepackets, $\rho_\text{e/h}$
exhibits a non-zero concurrence on a finite interval of $D$ that stops
before $D$ gets close to zero and starts a little before $D_\text{opt}$. 
The lower limit of the interval where electron/hole entanglement can be
observed with the delocalized wavepackets comes from decay of
electron/hole coherences as $D\rightarrow 0$. 
Although these wavepackets tend to be quite good
in terms of their overlaps with $\mathcal{G}_S^{(e)}$, the coherence
is too small to get a non-zero entanglement. For the
truncated wavepacket, the concurrence is higher than for delocalized
wavepackets as $D\rightarrow 0$ as expected since it is only for small
$D$ that these two wavepackets become significantly different.

Although this numerical investigation shows that the mesoscopic
capacitor generates some electron/hole pair entanglement in the low
$D$ limit, it also shows the importance of using the proper single-particle
states to study quantum information theoretical quantities and also to probe
the many-body nature of the electronic fluid.  We will now sketch a general
strategy inspired by signal processing for analyzing electronic coherences
generated by a time-periodic source.

\subsection{Electronic atoms of signal}
\label{sec:QI:atoms-of-signal}

The above discussion was based on the use of empirical electron and
hole wavefunctions without thinking of any detection scheme. However,
for a $T$-periodic source, one could envision time-dependent
detectors ideally performing repeated
detections of a given single-particle template across various periods. 
Although this is still not demonstrated experimentally, it is important
to develop a signal processing framework for the
corresponding signals.

The idea, borrowed from M. Devoret's lectures at Coll\`ege de
France~\cite{Devoret:CDF:2008} consists in using a family of
single-particle wavefunctions well suited to $T$-periodicity. We thus
consider single-particle wavepackets $\varphi_{k,l}$, which we call
electronic atoms of signal or, adapting M.~Devoret's terminology, electronic wavelets
such that
\begin{subequations}
\label{eq:wavelets}
\begin{align}
\label{eq:wavelets:time-shift}
\varphi_{k,l}(t)&=\varphi_k(t-lT)\\
\label{eq:wavelets:orthogonality}
\langle
\varphi_{k',l'}|\varphi_{k,l}\rangle &= \delta_{k,k'}\delta_{l,l'}\,.
\end{align}
\end{subequations}
Families $\varphi_{k,l}$ obtained by translations in both the frequency
and time domains arising from a single wavepacket $\varphi_{0,0}$ are
called Gabor bases. 
An important result for Gabor bases is the
Balian-Low theorem which states that there is no orthogonal basis of
this type that is well localized both in the time and frequency domain.
In the signal processing community, discrete wavelets are
families satisfying \eqref{eq:wavelets} in which the parameter $k$ corresponds to a scaling:
$\varphi_{k,l}(t)=(v_F\tau_0)^{-1/2}s^{-k/2}\varphi((t/s^n-lT)/\tau_0)$. 

A famous example of electronic atoms of signal has been introduced by
Th.~Martin and
R.~Landauer~\cite{Martin:1992-1}. These wavepackets with energy bandwidth
$hf$ ($f=1/T$) are defined by:
\begin{equation}
\varphi_{n,l=0}(t)=\frac{1}{\sqrt{v_FT}}\,\frac{\sin{(\pi ft)}}{\pi
	ft}\,\me^{-\mi\omega_nt}
\end{equation}
where $\omega_n=\pi f(2n+1/2)$. They are called Shannon
wavelets in the signal processing community. Remarquably,
when a voltage drive of period $T$ is applied, the single-particle
state $\varphi_{n,l}$ is scattered among the $\varphi_{n',l}$ with the
same time slot index $l$~\cite{Dubois:2013-1}.

More recently, M. Moskalets has
found a family of mutually orthogonal electronic
wavepackets $(\varphi_l)_{l\in\mathbb{Z}}$
respectively time-shifted by $lT$ and such that a train of single-electron
Leviton excitations is represented as the infinite Slater
determinant of these $\varphi_l$ on top of the Fermi
sea~\cite{Moskalets:2015-1}. It is then tempting to conjecture 
that there exists a family of
electronic atoms of signal $\varphi_{n,l}$ satisfying the orthogonality condition
\eqref{eq:wavelets:orthogonality} such that a train of charge $n$
Levitons is obtained as the infinite Slater determinant formed by the
$\varphi_{k,l}$ for $l\in \mathbb{Z}$ and $k=1,\ldots,n$ on top of the
Fermi sea.

Electronic atoms of signal can be used to model repeated detection 
of $\varphi_k$ by considering the
overlap of the single-electron coherence of a $T$-periodic source $S$
with a train of $N$ wavepackets $\varphi_{k,l}$ where $k$ is fixed and
$l=1,\ldots ,N$. 
This overlap represents the average cumulated signal after 
$N$ successive detections of 
the wavepacket $\varphi_k$ shifted by multiples of $T$. Due to the orthogonality
of the $\varphi_{k,l}$ for different values of $l$, these time-shifted
wavepackets are perfectly distinguishable: the corresponding number
operators $N[\varphi_{k,l}]$ and $N[\varphi_{k,l'}]$ commute. 
Using the $T$ periodicity of
$\mathcal{G}^{(e)}_S$, this overlap scales as $N$ with a prefactor
$\bar{n}[\varphi_k]$ defined by Eq. \eqref{eq:QI:population}.
The same reasoning can be extended to two-particle detection. We then
consider $\varphi_a$ and $\varphi_b$  for $a\neq b$ and we perform a
repeated detection of the electron pair. The associated wavefunctions
are the normalized Slater determinants 
$\varphi_{(a,b;l)}(t_1,t_2)$ built from $\varphi_{a,l}$ and
$\varphi_{b,l}$ for $l=1,\ldots ,N$. Exactly as in the single-particle
detection, due to orthogonality of the
single-particle wavefunctions for $l\neq l'$, the
two-particle states $\varphi_{(a,b;l)}$ are orthogonal for different
values of $l$. The cumulated average signal for $N$
periods is then the
overlap of $\mathcal{G}^{(2e)}$ with the sum over $l$ of 
$\varphi_{(a,b;l)}(t_1^+,t_2^+)\,\varphi_{(a,b;l)}(t_1^-,t_2^-)^*$.
Once again this overlap scales as $N$ with a prefactor
$\bar{n}[\varphi_a,\varphi_b]$ equal to
\begin{equation}
v_F^4\int_{\mathbb{R}^4}
\varphi_{(a,b)}(\mathbf{t_+})^*\varphi_{(a,b)}(\mathbf{t}_-)
\mathcal{G}^{(2e)}_S(\mathbf{t}_+|\mathbf{t}_-)\,\md^2\mathbf{t}_+\,\md^2\mathbf{t}_-\,.
\end{equation}
Throught repeated detections of the same
wavefunction in different time slots,
electronic atoms of signal thus enable us to
define effective density matrices in the basis of occupation number for
orthogonal single-particle levels.

To illustrate this point, Figs.~\ref{fig:source-LPA:ne-etc} and
\ref{fig:source-LPA:concurrence} also present the same
results as before but for Martin Landauer electronic atoms of signals
associated with the period $T/2$ so that $\varphi_e$ is centered in
energy at
$\omega_e$ and in time on the first half period and $\varphi_h$ is
centered in energy at $\omega_h$ and in time on the second half period.
As can be seen from the behavior of $\bar{n}_e$,
these atoms of signals are not well suited close to
$D_\text{opt}$. This was expected since their energy width $2hf$ is less
than the natural width of the emitted single-electron excitation. 
However, the electron/hole coherence $\xi$ is closer to the Cauchy-Schwarz
bound than the truncated and delocalized wavepackets. As a result, the
non-vanishing interval for the concurrence is indeed concentrated at lower
values than for the truncated and delocalized wavepackets. 

What we are doing here is to analyze the fermionic analogue of
inter-mode entanglement in quantum optics. Considering various
families of single-particle state amounts to considering various pairs
of modes. Naturally, quantum
information quantities measuring this entanglement depend on the modes
considered.
This raises the question of the best description of
a quantum signal such as single-electron coherence in terms of single-particle
wavefunctions. 
What are the guidelines for such a choice? Although we do not
have a definitive answer to this very general question, we think there are two ways
of addressing it. 

First of all, the experimental setup may impose us a choice. For example,
Moskalets wavefunctions and their conjectural
generalization are the natural choice when discussing 
an HOM experiment with a source emitting a periodic 
train of Leviton excitations in one of the incomming channels.
Next, when starting from computational or experimental data
on single-electron coherence, the
problem is to determine atoms of signals or other wavepackets giving
the ``simplest" description of single-electron coherence.  
Although
answering this question goes beyond the scope of the
present paper, methods inspired from signal processing and exploiting
the $T$-periodicity of single-electron coherence are certainly worth
investigating. 

\section{Conclusion and perspectives}
\label{sec:conclusion}

In this paper, we have discussed the recent developments in electron
quantum optics from a signal processing perspective. Although 
part of the material presented here forms a review of our recent works in
electron quantum
optics~\cite{Bocquillon:2014-1,Thibierge:2016-1,Marguerite:2016-2,Marguerite:2016-1,Ferraro:2013-1}, our
discussion aims at showing new interesting perspectives for future work
as well as connections to other topics such as quantum optics 
of current noise or the study of 
quantum information quantities in electronic systems.

From that perspective, 
our first main message is
that single- and two-particle interferometry experiments can be
interpreted in the signal processing language as analog operations
converting ``quantum signals'' such as single- and two-electron
coherences into experimentally observable quantities which are zero- 
or finite-frequency average current and current correlations. In particular, we
have reviewed how the
HOM experiment encodes the overlap of single-electron coherences within
low-frequency current noise, thus giving more substance to Landauer's aphorism
``The noise is the signal''~\cite{Landauer:1998-1}. We have also shown 
that the signal
processing point of view is relevant beyond the analysis of HOM
experiments. For example, it also unifies the
direct probing of intrinsic two-electron coherence
within a given edge channel by means of generalized Franson
interferometry experiments.

Our approach thus suggests that, although the demonstration of a Franson interferometer in
electron quantum optics represents a strong challenge due to interaction
effects within Mach-Zehnder interferometers, it might be easier to
probe two-electron coherences with a 
simple Hanbury Brown and Twiss interferometer by measuring correlations
between finite-frequency currents. Although this would require
sophisticated homodyning to bring the finite-frequency components of
both outgoing currents in the same low-frequency band, 
the measurement stage would be rather immune to
interaction effects. 
Interestingly, this would also establish a bridge between electron
quantum optics and the recently developped study of quantum
properties of the radiation emitted by a quantum
conductor~\cite{Grimsmo:2016-1}.

Our second message is that signal processing concepts and techniques are
useful not only for interpreting electron quantum optics experimental
results but also to gain a deeper understanding of the electronic many-body state
in these experiments. 

For this purpose, we have
transposed the concept of atoms of signal to electron quantum
optics and shown that it can be used to discuss quantum information
theory quantities in electron quantum optics experiments. We have
illustrated this idea by discussing the electron/hole entanglement
generated by the mesoscopic capacitor. However, our discussion is
relevant to the study of entanglement in more general setups
(see for example~\cite{Hofer:2016-2} in this volume and
\cite{Thomas:2016-1,Dasenbrook:2016-1} 
as well as
\cite{Samuelsson:2004-1,Chtchelkatchev:2002-1,Samuelsson:2003-1,Samuelsson:2005-1,Fazio:2006-1,Giovannetti:2007-1,Sherkunov:2012-1}
for previous works).
By relying only on the use of electron quantum optics coherences,
our framework enables discussing Coulomb interaction effects in a very
natural way. 
Second, we think that the pioneering work by
M.~Moskalets on Leviton trains~\cite{Moskalets:2015-1} raises the question
of the ``simplest
representation'' of quantum signals such 
as single-electron coherences emitted by various electronic
sources.
We think that combining the arsenal of theoretical techniques
(analytical as well as numerical) with signal processing concepts may
lead to progresses on this basic question. 

Last but not least, the rapid development of experiments in electron
quantum optics and microwave quantum optics suggests that the
perspectives for investigating all these questions on the experimental side
are very promising.

\begin{acknowledgement}
We warmly thank J.~Splettstoesser and R.~Haug for setting up this
special volume.
We acknowledge useful discussions with P.~Borgnat 
and R.~Menu.
This work was supported by the ANR grants ``1shot reloaded" (Grant No.
ANR-14-CE32-0017) and by the ERC consolidator grant ``EQuO" (Grant No.
648236).
\end{acknowledgement}

%\bibliographystyle{../src/pss}
%\bibliography{../src/biblio/bigbib}

\providecommand{\WileyBibTextsc}{}
\let\textsc\WileyBibTextsc
\providecommand{\othercit}{}
\providecommand{\jr}[1]{#1}
\providecommand{\etal}{~et~al.}

\end{document}